\documentclass[prd,aps,preprintnumbers,floats,floatfix,superscriptaddress,preprintnumbers,
showpacs,eqsecnum,nofootinbib,twocolumn]{revtex4-2}
\usepackage[utf8]{inputenc}
\usepackage{latexsym,array,theorem,mathrsfs,bm,float}
\usepackage{psfrag}
\usepackage{amsfonts,amsmath,amssymb,latexsym,array,afterpage,
theorem,mathrsfs,bm,float,epsfig,color,graphicx,tabularx,here,multirow}
\usepackage[linktocpage=true]{hyperref}

\newcommand{\nn}{\nonumber \\}
\newcommand{\Mpl}{M_{\rm Pl}}
\newcommand{\D}{{\rm d}}
\newcommand{\Lcal}{\mathcal{L}}

\newcommand{\be}{\begin{eqnarray}}
\newcommand{\ee}{\end{eqnarray}}

\newcommand{\bem}{\begin{bmatrix}}
\newcommand{\eem}{\end{bmatrix}}

\newcommand{\mn}{{\mu \nu}}

\newcommand{\mR}{\mathcal{R}}

\newcommand{\homega}{\hat{\omega}}

\begin{document}
\baselineskip=12pt

\preprint{YITP-22-155}
\title{
Highly compact Proca stars with quartic self-interactions
}
\author{Katsuki Aoki}
\affiliation{
Center for Gravitational Physics and Quantum Information, Yukawa Institute for Theoretical Physics, Kyoto University, 606-8502, Kyoto, Japan
}

\author{Masato Minamitsuji}
\affiliation{Centro de Astrof\'{\i}sica e Gravita\c c\~ao  - CENTRA, Departamento de F\'{\i}sica, Instituto Superior T\'ecnico - IST, Universidade de Lisboa - UL, Av. Rovisco Pais 1, 1049-001 Lisboa, Portugal}


\date{\today}

\begin{abstract}
We study self-gravitating bound states of a complex vector field, known as Proca stars, with a new type of quartic-order self-interaction which does not exist in the case of either a complex scalar field or a real vector field. Depending on the sign of the coupling constant, this quartic self-interaction can yield a distinct feature of Proca stars from the previously investigated self-interaction of the vector field. We find that self-gravitating solutions can be so compact that the photon sphere could form. However, we also show that the self-interaction gives rise to a ghost instability for the stars whose compactness is close that for the formation of a photon sphere, which might invalidate the formation of the photon sphere.
\end{abstract}

\maketitle

\section{Introduction}
\label{sec1}

While current experimental and observational data are consistent with the predictions of General Relativity (GR) \cite{Clifton:2011jh,Will:2014kxa},
future observations of gravitational waves and those associated with the strong-field regime
will provide us new opportunities to test the validity of GR \cite{Berti:2018cxi,Berti:2018vdi,Barack:2018yly}.
The possibility of hypothetical horizonless compact objects
\cite{Cardoso:2017cqb,Cardoso:2019rvt,Berti:2018cxi,Berti:2018vdi,Barack:2018yly}
would also provide us to test the existence of the new particle sectors in the strong-field regime.
Boson stars
are the representative candidates of such compact objects
\cite{Jetzer:1991jr,Schunck:2003kk,Visinelli:2021uve,Liebling:2012fv}
and are characterized by
the Arnowitt-Deser-Misner (ADM) mass and Noether charge
associated with the global $U(1)$ charge \cite{Kaup:1968zz,Ruffini:1969qy,Friedberg:1986tp}.
With an increase of the scalar amplitude,
the ADM mass and Noether charge increase
and 
the solutions below reaching the maximal values of them
are dynamically stable 
\cite{Gleiser:1988rq,Gleiser:1988ih,Hawley:2000dt}.
Boson star  solutions also exist
in the presence of self-interacting potentials \cite{Schunck:2003kk,Guerra:2019srj}.

Boson star solutions can be naturally extended to the complex Proca field, which are known as Proca stars
\cite{Brito:2015pxa,Brihaye:2017inn,Garcia:2016ldc,Minamitsuji:2018kof,Herdeiro:2020jzx,Cardoso:2021ehg,Zhang:2021xxa} (see also Ref.~\cite{Jain:2022kwq} for a Yang-Mills field and Refs. \cite{Aoki:2017ixz,Brito:2020lup,Jain:2021pnk} for a spin-2 field).
In the massive Proca theories, 
the properties of 
Proca stars are very similar to those of scalar boson stars~\cite{Brito:2015pxa},
and have been extensively applied to
astrophysics and gravitational-wave physics
\cite{Sanchis-Gual:2018oui,DiGiovanni:2020ror,Bustillo:2020syj,Herdeiro:2021lwl,Rosa:2022tfv,Rosa:2022toh}.
In the presence of the self-interaction potential of the complex Proca field $V(\bar{A}^\mu A_{\mu})$,
however, the situation is drastically modified.
Proca star solutions
cease to exist
for the central Proca amplitude at a critical point \cite{Minamitsuji:2018kof,Herdeiro:2020jzx,Cardoso:2021ehg}.
The problem arises when the first radial derivative of the radial component 
of the Proca field diverges at a certain radius,
beyond which one cannot integrate the field equations numerically.
We have recently verified that 
the appearance of a singular point at a finite amplitude 
can be interpreted as the onset of a gradient instability
at the background level \cite{Aoki:2022woy}.
An independent but conceptually related problem is a ghost instability of 
a self-interacting Proca field \cite{Clough:2022ygm,Coates:2022qia,Mou:2022hqb}
(see also Ref.~\cite{Coates:2022nif}).
Here, ``ghost'' and ``gradient'' instabilities are associated with the wrong signs of the kinetic and gradient terms
in the Lagrangian, respectively.
They grow arbitrarily fast if they continue existing in arbitrarily high-energy/-momentum scales,
and the presence of these instabilities invalidates the perturbation theory within an infinitesimally short timescale.
At the onset of a ghost or gradient instability,
the hyperbolicity of equations of motion is lost.
References~\cite{Clough:2022ygm,Coates:2022qia,Mou:2022hqb}
perform numerical simulations of the time evolution
of a self-interacting real Proca field in different backgrounds,
and show
that the time derivative of the temporal component of the Proca field
diverges at a certain moment of time,
beyond which one cannot follow the time evolution.
The problem of a ghost instability is expected to be generic to the self-interacting Proca sector
and independent of background geometries.

A ghost or gradient instability is indeed a pathology of the theory,
if the self-interacting Proca field is a fundamental field.
However,
the self-interacting Proca theory
could appear as  a low-energy effective description of a more fundamental theory,
and then ghost and gradient instability problems may be cured,
once ultraviolet (UV) physics such as the dynamics of heavy fields is properly taken into consideration \cite{Aoki:2022woy} (see also Refs.~\cite{Zhang:2021xxa,Jain:2022kwq,Fukuda:2019ewf,East:2022ppo}).
Reference~\cite{Aoki:2022woy} proposed a simple partial UV completion model of the self-interacting Proca theory
 with a new heavy scalar field.
From the effective field theory (EFT) viewpoint,
the onset of a gradient or ghost instability indicates a breakdown of the self-interacting Proca field as an EFT. 
It has been demonstrated that 
Proca star solutions,
more precisely Proca-scalar star solutions as there is a scalar field in addition to the Proca field, 
continue to exist even beyond the critical point at which the EFT suffers from a gradient instability,
and a small deviation is enough to regularize the singularity in the EFT.
Several physical properties of the Proca-scalar star solutions (and nongravitating solutions) were studied 
by the authors of Ref.~\cite{Herdeiro:2023lze}
\footnote{Reference~\cite{Herdeiro:2023lze} uses an Abelian-Higgs-like model as a partial UV completion of the quartic-order self-interaction. Note, however, that their model is different from the Abelian-Higgs model (see e.g., Refs.~\cite{Fukuda:2019ewf,East:2022ppo}) because the vector field is complex rather than real. For instance, the field space of the scalar sector is no longer flat, differently from the Abelian-Higgs model (see e.g.,~Eq.~(4.10) of Ref.~\cite{Aoki:2022woy}).}.
Then, it was shown that the maximal mass and compactness of 
Proca-scalar stars
cannot exceed those of Proca stars
in the pure Einstein-massive complex Proca theory
which is recovered in a certain limit
from the UV theory,
and the photon spheres and the innermost stable circular orbits
cannot be formed in the stable branch.

Although Ref.~\cite{Aoki:2022woy} focused on the quartic-order self-interaction $(\bar{A}^\mu A_{\mu})^2$,
analysis can be naturally extended to a more general self-interaction
including higher-order terms of ${\bar A}^\mu A_\mu$.
The issue on the existence of Proca stars provided one of the simplest setups
that demonstrate a partial UV completion to cure the pathology of the self-interacting Proca field,
in the sense that the system is given by a set of ordinary differential equations.
While the problems of perturbations of the self-interacting Proca field on a nontrivial background
\cite{Clough:2022ygm,Coates:2022qia,Mou:2022hqb}
are formulated by a set of partial differential equations depending on both the space and time,
they can be similarly resolved.
Starting from an initial condition with a small amplitude where the EFT is valid,
the system may evolve into a large amplitude of the vector field.
Once UV physics is taken into account,
the heavy mode should be excited before the breakdown of the EFT.

In the present paper, we will focus on another aspect of the self-interacting complex Proca
field on the Proca star background while paying attention to UV physics.
We will study properties of Proca stars
in the presence of the new self-interaction of a complex Proca field $A_{\mu}$,
which includes the scalar product ${\bar A}^\mu {\bar A}^\nu A_\mu A_\nu$
[see Eq. \eqref{quarticpotential} below]
and vanishes in the limit of a real vector field.
We will show that 
this new self-interaction could realize distinctive features of Proca stars,
in comparison with those in the presence of the higher-order powers of ${\bar A}^\mu A_\mu$
explored in the literature \cite{Minamitsuji:2018kof,Herdeiro:2020jzx,Cardoso:2021ehg}.
We expect that, as in the case of the other theories,
a possible ghost or gradient instability arises from the new self-interaction and 
they could also be cured by UV physics.
Hence, we will concentrate on the properties of Proca stars
in the regime of the EFT in the present paper.
Yet, the underlying assumptions about UV physics lead to nontrivial consequences on Proca stars,
 as we will discuss below.

This article is organized as follows.
In Sec. \ref{sec2},
we introduce the Einstein-complex Proca theory with the new self-interaction of 
a complex Proca field, which vanishes in the limit of a real Proca field.
In Sec. \ref{sec3},
we provide the basic equations to construct Proca star solutions.
In Sec. \ref{sec4},
we discuss the properties of the Proca star solutions in our theory,
especially focusing on the maximal compactness and 
the existence of a ghost or gradient instability.
The last section~\ref{sec5}
is devoted to giving a brief summary and conclusion.

\section{Self-interacting complex Proca fields}
\label{sec2}

Before initiating our analysis,
it is instructive to mention differences among a complex scalar, a real vector field, and a complex vector field,
to see how the vectorial and complex natures lead to the new interactions
absent in the cases of the scalar field and the real vector field. 
Throughout the present paper, we will ignore derivative interactions as they may be suppressed in low energies.

We first consider the following Lagrangian density of a complex scalar field $\Phi$,
\begin{align}
\label{complex_scalar}
\Lcal_{\Phi} = - \frac{1}{2} \nabla^{\mu} \bar{\Phi} \nabla_{\mu} \Phi  - V_{\Phi} (\bar{\Phi}\Phi)
\end{align}
where the overbar represents the complex conjugate, $V_{\Phi}$ is the potential,
and $\nabla_\mu$ denotes the covariant derivative associated with the spacetime metric $g_\mn$.
Note that here and in the rest the spacetime indices are raised by the inverse metric tenor $g^\mn$,
such as $\nabla^\mu=g^\mn \nabla_\nu$,
and lowered by the metric tenor $g_\mn$.
Since the Lagrangian density \eqref{complex_scalar} must be real, 
the potential is a function of the modulus $\bar{\Phi}\Phi$. 
For instance, the potential up to the quartic order is given by 
$V_{\Phi} (\bar{\Phi}\Phi)=\frac{\mu^2_{\Phi} }{2} \bar{\Phi} \Phi + \frac{\lambda_{\Phi} }{4} (\bar{\Phi} \Phi)^2$,
where $\mu^2_{\Phi}$ is the mass squared around the vacuum $\langle \Phi \rangle =0$ 
which we shall assume positive. The coupling constant $\lambda_{\Phi} $ can be either positive or negative; 
a positive $\lambda_{\Phi}$ represents a repulsive self-interaction,
while a negative $\lambda_{\Phi}$ leads to an attractive self-interaction.

We then consider a real vector field $B_{\mu}$ with the Lagrangian
\begin{align}
\Lcal_B = -\frac{1}{4} B^{\mu\nu} B_{\mu\nu} - V_B (B^{\mu}B_{\mu})
\label{Lag:realvector}
\end{align}
where $B_{\mu\nu}
:= \partial_{\mu}B_{\nu}-\partial_{\nu}B_{\mu}$. 
The potential $V_B (B^{\mu}B_{\mu}) $ is a function of $B^{\mu}B_{\mu}$ since $V_B$ should be a Lorentz-invariant scalar. 
Let us write it as $V_B (B^{\mu}B_{\mu})=\frac{\mu^2_B }{2} B^{\mu}B_{\mu}+ \frac{\lambda_B }{4} (B^{\mu}B_{\mu})^2$
up to the quartic order, which has a structure similar to the scalar potential $V_{\Phi}$.
Note that a self-interacting massive vector field is not UV complete on its own. 
Hence, Eq.~\eqref{Lag:realvector} should be regarded as a low-energy EFT,
and a UV completion is needed when the theory loses its validity. The simplest example is the Higgs mechanism in which the self-interaction $(B^{\mu}B_{\mu})^2$ arises as a result of integrating out the Higgs field.
See, e.g., Refs.~\cite{Zhang:2021xxa, Jain:2022kwq, Aoki:2022woy} in the context of boson stars. 
One may then notice that the sign of $\lambda_B$ is necessarily negative~\cite{Fukuda:2019ewf,East:2022ppo}
because it is determined by the squares of the coupling constant and the mass. 
The negative sign is not specific to the Higgs mechanism; it has to be negative for any (nongravitational) UV completion
under the fundamental assumptions, namely unitarity, Poincar\'{e} invariance, 
causality, and locality, and the resultant bounds on low-energy EFTs 
are called positivity bounds~\cite{Adams:2006sv}.
It is, however, worth mentioning that the positivity bounds 
in gravitational systems are still subject to discussions and, especially, 
the sign of the Planck suppressed operators may depend on details of quantum gravity~\cite{Hamada:2018dde, Alberte:2020jsk, Tokuda:2020mlf, Herrero-Valea:2020wxz, Caron-Huot:2021rmr, Alberte:2021dnj, Herrero-Valea:2022lfd}.
In the context of boson stars, the self-interactions are often supposed to be Planck suppressed [see Eq.~\eqref{dimless} later], so it would be also interesting to investigate the positive value of $\lambda_1$ even though its UV completion is not known.

Now, we turn to our main focus, a complex vector field $A_{\mu}$
whose Lagrangian density is
\begin{align}
\label{proca_lagrangian}
\Lcal = -\frac{1}{4}\bar{F}^{\mu\nu}F_{\mu\nu} - V
\end{align}
where $F_{\mu\nu} := \partial_{\mu}A_{\nu}-\partial_{\nu}A_{\mu}$,
and 
the overbars again represent the complex conjugate.
In the literature on Proca stars \cite{Brito:2015pxa,Brihaye:2017inn,Garcia:2016ldc,Minamitsuji:2018kof,Herdeiro:2020jzx,Cardoso:2021ehg,Zhang:2021xxa,Jain:2022kwq}, 
the self-interacting potential $V$ is often assumed to be a function of 
\be
\label{defy}
Y:=\bar{A}^{\mu}A_{\mu}
\ee
just like the cases of a complex scalar field and a real vector field. 
However, there is the missing scalar invariant given by 
\be
\label{defz}
Z:=\bar{A}^{\mu}\bar{A}_{\mu}A^{\nu}A_{\nu}
\ee
which is independent from $(\bar{A}^{\mu}A_{\mu})^2$,
since $A_{\mu}$ has a spacetime index and is complex. 
Therefore, the most general potential up to the quartic order of $A_\mu$ and ${\bar A}_\mu$
is given by
\begin{align}
V&=\frac{\mu^2}{2}Y + \frac{\lambda_1}{4}Y^2
 + \frac{\lambda_2}{4}(Z-Y^2)
\,,
\label{quarticpotential}
\end{align}
where
$\mu$ is the mass of the Proca field,
and 
$\lambda_1$ and $\lambda_2$ are dimensionless self-coupling constants.
The last term in Eq. \eqref{quarticpotential} vanishes 
when $A_{\mu}$ is real, i.e.,~$\bar{A}_{\mu}=A_{\mu}$. 
As shown in Appendix~\ref{sec:positivity}, the (nongravitational) positivity bounds conclude
\begin{align}
\lambda_1<0\,, \quad \lambda_2<0
\,.
\label{positivity_lambda}
\end{align}
The appearance of the second-type self-interaction from a Yang-Mills theory is well known,
 e.g., the self-interaction of the $W$ boson, although the theory contains other vector fields in addition to the complex Proca field in this case.
Nonrelativistic bosonic bound states in a Yang-Mills-Higgs system were found in Ref.~\cite{Jain:2022kwq} in which the self-interactions of the $SU(2)$ Yang-Mills field, i.e., three real vector fields, were studied.
It should be, nevertheless, stressed that the bounds \eqref{positivity_lambda} are universal under the above-mentioned assumptions,
regardless of whether $A_{\mu}$ is a part of a Yang-Mills field.

Therefore, there are two differences between the complex scalar field (scalar boson stars) and the complex vector field (Proca stars): the presence of the new self-interaction and the close connection between the sign of the coupling constants and UV physics. If the complex Proca field has the standard UV completion, the coupling constants must satisfy the inequalities \eqref{positivity_lambda} regardless of the details of UV completion. On the other hand, $\lambda_1>0$ or $\lambda_2>0$ requires unknown UV completion. The sign of the quartic order self-interaction of the Proca field has significant information about UV physics, and from the point of view of a bottom-up approach, the sign is determined through observations.

Having understood these distinctions, we study the properties of relativistic Proca star solutions
in the Einstein-Proca theory
\be
\label{eft}
S=
\int d^4x
\sqrt{-g}
\left[
\frac{\Mpl^2}{2}R
 -\frac{1}{4}\bar{F}^{\mu\nu}F_{\mu\nu} - V
\right],
\ee
where the Einstein-Hilbert term $(\Mpl^2/2)R$ is added
as the gravitational kinetic term 
to the self-interacting complex Proca field \eqref{proca_lagrangian}.
Here, 
$\Mpl$ is the reduced Planck mass,
$R$ represents the Ricci scalar associated with the metric $g_{\mu\nu}$,
and 
$V$ is given by Eq.~\eqref{quarticpotential}.

Varying the action \eqref{eft} with respect to $g_{\mu\nu}$,
we obtain the gravitational equations of motion
\be
\label{grav_eom}
\Mpl^2 G_{\mu\nu}=T_{\mu\nu},
\ee
where we have defined the energy-momentum tensor 
of the complex Proca field
\be
\label{tmn}
T_{\mu\nu}
&=&
F_{(\mu}{}^\alpha {\bar F}_{\nu)\alpha}
-
\frac{1}{4}
g_{\mu\nu}
{\bar F}^{\rho\sigma}{\bar F}_{\rho\sigma}
-V g_{\mu\nu}
\nonumber\\
&+&
\left[
\mu^2 +(\lambda_1-\lambda_2)Y
\right]
A_{(\mu}{\bar A}_{\nu)}
\nonumber\\
&+&
\frac{\lambda_2}{2}
\left(
{\bar A}_\mu {\bar A}_\nu
A^\rho A_\rho
+
A_\mu A_\nu
{\bar A}^\rho {\bar A}_\rho
\right).
\ee
Varying the action \eqref{eft} with respect to ${\bar A}^\nu$,
we obtain the Proca equation of motion
\be
\label{eomProca}
&&
\nabla_\mu F^{\mu\nu}
-
\left[
\left(
\mu^2 +\lambda_1A_\rho {\bar A}^\rho
\right)
A^\nu
\right.
\nonumber\\
&&
\left.
+
\lambda_2
 A^\rho
\left(
- {\bar A}_\rho A^\nu
+ A_\rho {\bar A}^\nu
\right)
\right]=0.
\ee
Acting $\nabla_\nu$ on Eq. \eqref{eomProca},
we obtain the constraint condition
\be
&&
\nabla_\nu
\left[
\left(
\mu^2 +\lambda_1A_\rho {\bar A}^\rho
\right)
A^\nu
\right.
\nonumber\\
&&
\left.
+
\lambda_2
A^\rho
\left(
-{\bar A}_\rho A^\nu
+
A_\rho {\bar A}^\nu
\right)
\right]
=0.
\nonumber\\
\label{eomdiv}
\ee
There exists the global $U(1)$ symmetry under the transformation $A_\mu\to e^{i\alpha}A_\mu$
with a constant $\alpha$
and the associated Noether current is given by 
\be
\label{noether}
j^\mu
=
\frac{i}{2}
\left(
{\bar F}^{\mu\nu} A_\nu
-F^{\mu\nu}{\bar A}_\nu
\right),
\ee
which satisfies the local conservation law $\nabla_\mu j^\mu=0$.

\section{Proca star solutions}
\label{sec3}

\subsection{Background equations}
\label{sec3a}

In this section,
we present the basic equations to determine the structure of Proca star solutions
in the theory \eqref{eft} with Eq.~\eqref{quarticpotential}.
We assume the following ansatz for the metric and Proca fields,
\be
\label{metric_ansatz}
g_{\mu\nu}dx^\mu dx^\nu
&=&
-\sigma(r)^2
\left(
1-\frac{2m(r)}{r}
\right)
d\hat{t}^2
\nonumber
\\
&&
+
\left(
1-\frac{2m(r)}{r}
\right)^{-1}
dr^2
+
r^2
d\Omega^2,
\\
\label{Proca_ansatz}
A_\mu dx^\mu
&=&
e^{-i{\hat\omega} \hat{t}}
\left(
a_0(r)dt
+
i a_1 (r)dr
\right),
\ee
where
$\hat{t}$ and $r$ are the temporal and radial coordinates,
respectively;
$\sigma(r)$, $m(r)$, $a_0(r)$, and $a_1(r)$
are the real functions of $r$;
and 
$\homega$ is the frequency 
which is determined under suitable boundary conditions.
For a Proca star solution, 
the frequency ${\hat\omega}$ is real and positive,
so the Proca field is stationary and neither grows nor decays in time.
For a time-dependent complex Proca field,
each component of the energy-momentum tensor \eqref{tmn}
would in general become time dependent.
However,
we stress that 
the ansatz for the Proca field \eqref{Proca_ansatz}
is compatible with the staticity and spherical symmetry of the spacetime \eqref{metric_ansatz}
because
in each component of the energy-momentum tensor \eqref{tmn}
the explicit time dependence $e^{-i{\hat\omega} \hat{t}}$
in $A_\mu$ and its derivative
is canceled out
by its complex conjugate $e^{i{\hat\omega} \hat{t}}$
in ${\bar A}_\mu$ and its derivative for the ansatz Eq.~\eqref{Proca_ansatz}.
Hence, the components of the energy-momentum tensor \eqref{tmn}
do not contain the time dependence.

Substituting the above ansatz 
into the Einstein and Proca field equations,
Eqs.~\eqref{grav_eom} and \eqref{eomProca}
we find that $\sigma(r)$, $m(r)$, $a_0(r)$, and $a_1(r)$
obey the first-order ordinary differential equations
\be
\label{set_of_eqs}
\frac{dm}{dr}
&=&
F_m
\left[
m(r),\sigma(r),a_0(r),a_1(r);
r
\right],
\\
\label{set_of_eqs2}
\frac{d\sigma}{dr}
&=&
F_\sigma
\left[
m(r),\sigma(r),a_0(r),a_1(r);
r
\right],
\\
\label{set_of_eqs3}
\frac{da_0}{dr}
&=&
F_{0}
\left[
m(r),\sigma(r),a_0(r), a_1(r);
r
\right],
\\
\label{set_of_eqs4}
\frac{da_1}{dr}
&=&
F_1
\left[
m(r),\sigma(r),a_0(r),a_1(r);
r
\right],
\ee
where
\begin{widetext}
\be
F_m
&:=&
\frac{1}{8\Mpl^2}
\Bigg[
2r\mu^2 a_1^2 (r-2m)
+
4 a_1^2
\left(
r\mu^2+\lambda_1 a_1^2 (r-2m)
\right)
(r-2m)
\nonumber\\
&&
+
\lambda_1
a_1^4 
(r-2m)^2
-
\frac{2r^2\homega^2a_1^2}{\sigma^2}
-
\frac{2r^2 (\lambda_1-2\lambda_2)a_0^2a_1^2}
        {\sigma^2}
\nonumber\\
&&
-\frac{r^2}{\sigma^4}
\Big[
  \frac{3r^2\lambda_1 a_0^4}{(r-2m)^2}
-\frac{2r\mu^2 a_0^2\sigma^2}{r-2m}
-\frac{2a_1^2\sigma^2}{\homega^2}
\Big\{
\homega^2
+
(\lambda_1-2\lambda_2)
a_0^2
-
\frac{(r\mu^2+\lambda_1 a_1^2(r-2m)) (r-2m)\sigma^2}
        {r^2}
\Big\}^2
\Big]
\Bigg],
\\
F_\sigma
&:=&
-\frac{1}{2\Mpl^2}
\Big[
\frac{r^4\lambda_1a_0^4}
        {(r-2m)^3\sigma^3}
-\frac{r^3\mu^2 a_0^2} 
          {(r-2m)^2\sigma}
-a_1^2
\left(
r\mu^2
+
\lambda_1 a_1^2 (r-2m)
\right)
\sigma
\Big],
\\
F_{0}
&:=&
\frac{a_1}{\homega}
\left(
\homega^2
+(\lambda_1-2\lambda_2)a_0^2
-
\frac{\sigma^2
\big(
r\mu^2+\lambda_1 a_1^2 (r-2m)
\big)
(r-2m)}
       {r^2}
\right),
\ee
\end{widetext}
and $F_1$ is a complicated function of $(m,\sigma,a_0,a_1)$,
which contains 
\be
\label{calH}
{\cal H}(r)
&:=&
-
r^2
\left(\lambda_1-2\lambda_2\right)a_0^2
\nonumber\\
&+&
\left(
r\mu^2+3\lambda_1 a_1^2  (r-2m)
\right)
(r-2m)
\sigma^2,
\ee
in the denominator.
In integrating Eqs.~\eqref{set_of_eqs}--\eqref{set_of_eqs4},
in case one hits the point where ${\cal H}(r)=0$,
one cannot integrate them beyond this point,
and hence there is no Proca star solution.
As we will see later,
the appearance of the point ${\cal H}(r)=0$
can be interpreted as the onset of a gradient instability at the background level
\cite{Aoki:2022woy}.

\begin{figure}[h]
\unitlength=1.1mm
\begin{center}%
  \includegraphics[height=4.0cm,angle=0]{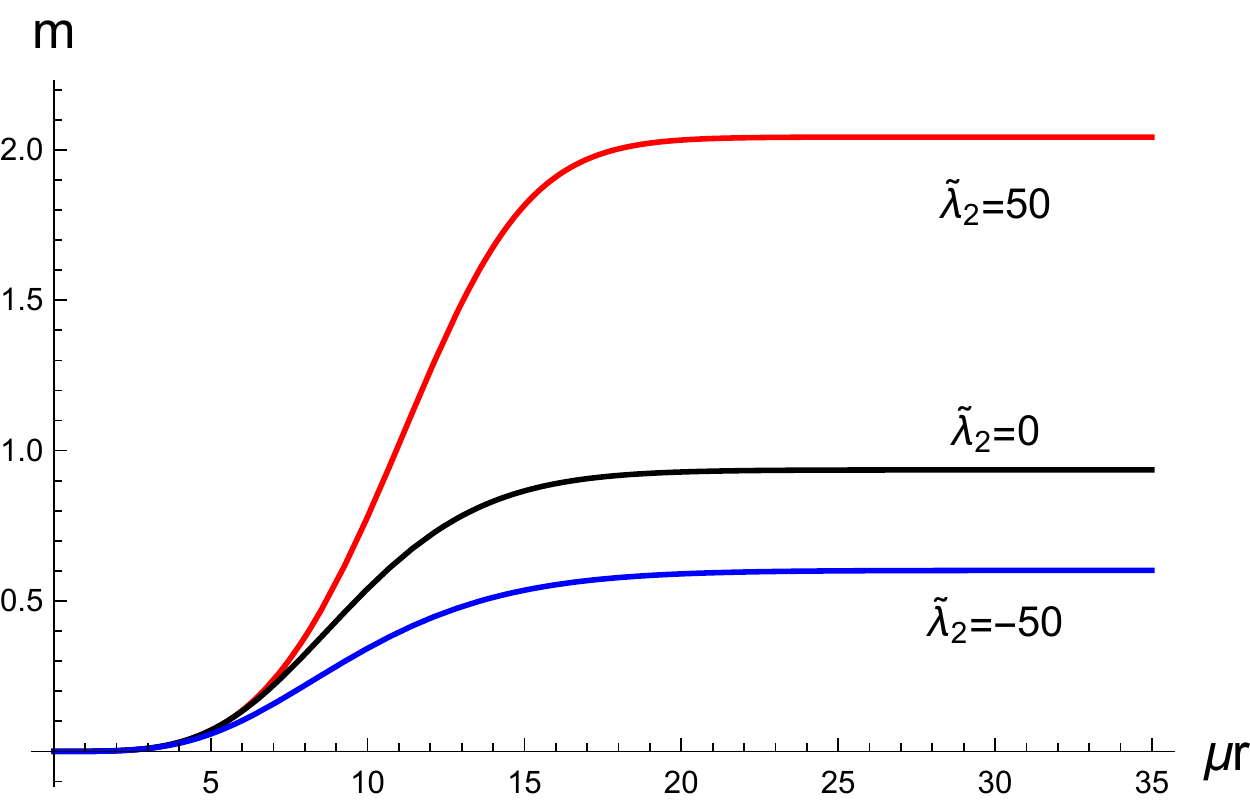}
  \includegraphics[height=4.0cm,angle=0]{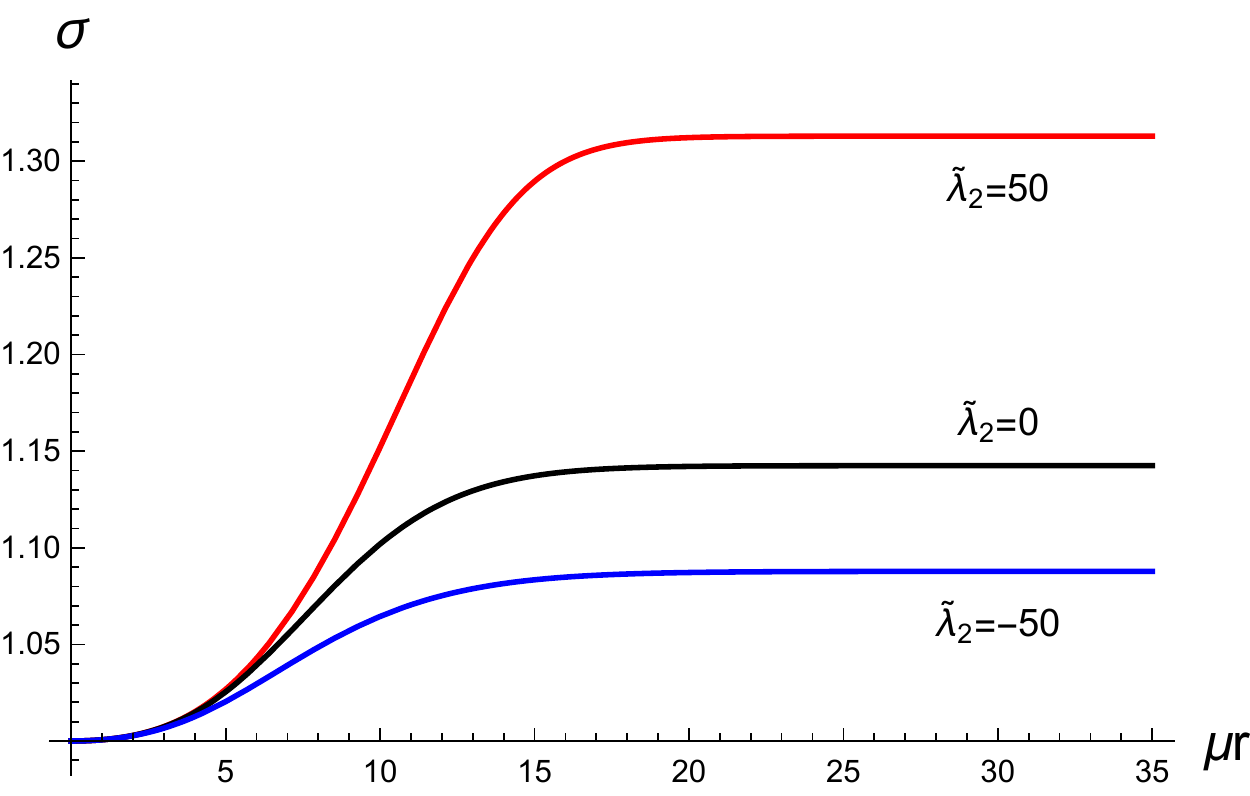}
  \includegraphics[height=4.0cm,angle=0]{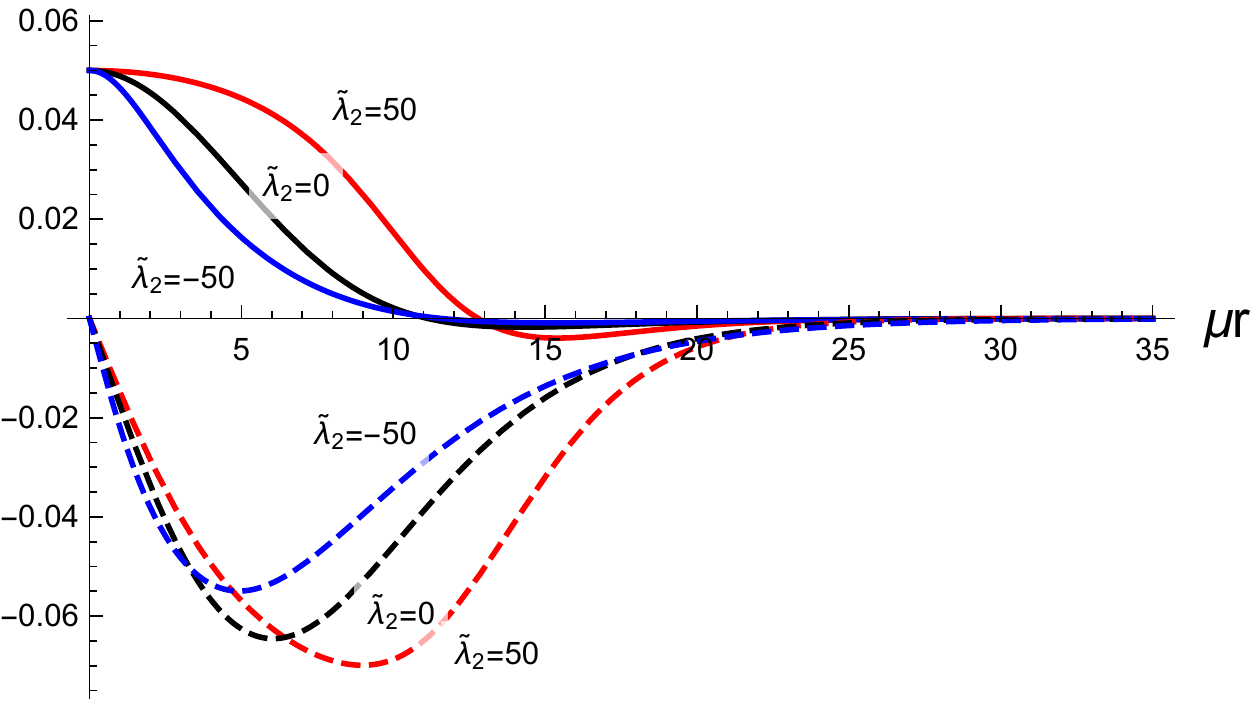}
\caption{
The profiles of 
the metric functions $m$ (top), $\sigma$ (middle),
and the Proca field (bottom)
for Proca star solutions in the ground state
are shown as the functions of $\mu r$
for $f_0=0.05$ with the specified coupling constants.
 In the bottom panel
the solid and dashed curves correspond to
the profiles
$a_0(r)$ and $a_1(r)$,
respectively.
Note that 
the dimensionless coupling constant ${\tilde \lambda}_2$
is defined in Eq. \eqref{dimless}.
}
  \label{fig_profile}
\end{center}
\end{figure}

Solving Eqs. \eqref{set_of_eqs}--\eqref{set_of_eqs4}
in the vicinity of the center $r=0$,
the regular solution can be obtained as 
\be
\label{bc1}
m(r)
&=&
\frac{
f_0^2
\left(
-3f_0^2\lambda_1
+2 \mu^2\sigma_0^2
\right)}
       {24\Mpl^2\sigma_0^4}
r^3
+{\cal O} (r^4),
\\
\label{bc2}
\sigma(r)
&=&
\sigma_0
+
\frac{
f_0^2
\left(
-f_0^2\lambda_1
+\mu^2\sigma_0^2
\right)}{4\Mpl^2\sigma_0^3}
r^2
+{\cal O} (r^4),
\\
a_0(r)
&=&
f_0 
\nonumber\\
&-&
\frac{f_0\left(f_0^2\lambda_1-\mu^2\sigma_0^2\right)
\left(
f_0^2\left(\lambda_1-2\lambda_2\right)
-\mu^2\sigma_0^2
+\homega^2
\right)
}
          {6f_0^2\left(\lambda_1-2\lambda_2\right)\sigma_0^2-6\mu^2\sigma_0^4}
r^2
\nonumber\\
&+&
{\cal O} (r^4),
\label{bc3}
\\
a_1(r)
&=&
-\frac{f_0\homega\left(f_0^2\lambda_1-\mu^2\sigma_0^2\right)}
          {3\sigma_0^2f_0^2\left(\lambda_1-2\lambda_2\right)-3\mu^2\sigma_0^4}
  r
+{\cal O} (r^3).
\label{bc4}
\ee
Under the regular boundary conditions \eqref{bc1}--\eqref{bc4} near the origin $r=0$,
we numerically integrate Eqs. \eqref{set_of_eqs}--\eqref{set_of_eqs4}
toward infinity.
For a given set of parameters,
if we choose ${\hat\omega}$ to be a proper value,
$m$ and $\sigma$ exponentially approach constant values, 
$m_\infty>0$ and $\sigma_\infty>\sigma_0>0$,
respectively,
and
$a_0$ and $a_1$ exponentially approach zero as $e^{-\sqrt{\mu^2-\omega^2}r}$,
where 
the proper frequency for the observer sitting at the infinity $r\to \infty$ is defined by
\begin{align}
\label{hatom}
{\omega}:=\frac{{\hat\omega}}{\sigma_\infty}.
\end{align}
The exponential falloff of the Proca field
in the large distance limit as $e^{-\sqrt{\mu^2-\omega^2}r}$
requires 
\be
\omega <\mu.
\ee
Thus, as $r\to \infty$,
the metric exponentially approaches the Schwarzschild form
\begin{align}
\label{metric}
ds^2
&\to 
-\sigma_\infty^2 
  \left(
   1-\frac{2m_\infty}{r}
  \right)d{\hat t}^2
\nonumber\\ 
&+  
\left(1-\frac{2m_\infty}{r}\right)^{-1}
dr^2
+r^2d\Omega_2^2,
\end{align}
where the proper time measured 
by the observer at $r=\infty$ is given by 
$t=\sigma_\infty {\hat t} $.
In the limit $f_0\to 0$, 
$\omega\to \mu$,
the Proca field profiles $a_0(r)$ and $a_1(r)$ trivially vanish,
and the Minkowski solution is obtained.
Note that 
there can be multiple solutions satisfying the same boundary conditions
near the origin and in the large distance region for discrete eigenvalues of $\omega$,
for a given set of the parameters.
In this paper, 
we will focus only on ground state solutions
obtained for the lowest eigenvalue of $\omega$,
where $a_0(r)$ and $a_1(r)$ have one and zero nodes, respectively (see Fig \ref{fig_profile}).

There are several conserved charges 
that characterize the properties of Proca stars.
The first is the ADM mass
\begin{align}
\label{adm}
M
:=
8\pi \Mpl^2 m_\infty,
\end{align}
which is determined by the asymptotic value of the mass function $m(r)$.
The second is 
the Noether charge associated with the global $U(1)$ symmetry,
which is given by 
integrating $j^{\hat t}$ in Eq. \eqref{noether}
over a constant-${\hat t}$ hypersurface
\be
Q
&:=&
\int_\Sigma d^3x \sqrt{-g} j^{\hat{t}}
\nonumber
\\
&=&
\int_0^\infty dr
\frac{4\pi r^2a_1 ({\hat \omega}a_1-a_0')}
       {\sigma},
\ee
where $\Sigma$ denotes a constant time hypersurface.
$Q$ physically describes the particle number of the Proca field,
and the Proca star is gravitationally bound,
when
\be
\label{bound}
\mu Q-M>0.
\ee

In the case of Proca stars as well as boson stars,
there is no unique definition of the radius,
and so we need to introduce an effective radius which characterizes 
the size of the distribution of the Proca field.
First, we define the radius ${\cal R}_{95}$ by
\be
\label{r95}
m({\cal R}_{95})
=
0.95\times m_\infty,
\ee
at which the mass function $m(r)$ reaches $95\%$ of the ADM mass
[see Eq.~\eqref{adm} for the relation of $M$ and $m_\infty$].
We also introduce the radius normalized by $Q$~\cite{Schunck:2003kk}
\be
\label{reff}
{\cal R}
&:=&
\frac{1}{Q}
\int_\Sigma d^3x 
\sqrt{-g}
 \left(r j^{\hat t}\right)
\nonumber\\
&=& 
\frac{4\pi}{Q}
 \int_0^\infty 
dr
\frac{
r^3a_1
\left(
 {\hat\omega} a_1
-a_0'
\right)}
{\sigma}.
\ee
While $\mR_{95}$ can be interpreted as
an astronomical surface of a Proca star within which the most energy of the Proca field is confined, 
$\mR$ can be interpreted as the expectation value of the radius
that characterizes the distribution of the Proca field.
We will numerically find ${\cal R}_{95}> {\cal R}$ in a typical Proca star solution.

By introducing the rescaled dimensionless quantities by
\be
\label{dimless}
&&
{\tilde \omega}
:=\frac{{\hat\omega}}{\mu},
\quad
\tilde{r}:=r \mu,
\quad
\tilde{m}:= \mu m,
\nonumber\\
&&
\tilde{a}_0 := \frac{a_0}{\Mpl},
\quad
\tilde{a}_1:= \frac{a_1}{\Mpl},
\quad
\tilde{\lambda}_{1,2} 
:=
\frac{\Mpl^2 \lambda_{1,2}}
        {\mu^2},
\ee
the evolution equations \eqref{set_of_eqs}--\eqref{set_of_eqs4}
can be rewritten into the form without
the dimensionful quantities $\mu$ and $\Mpl$.
Thus, without loss of generality,
for the numerical analysis, we set $\mu=\Mpl=1$, 
and if necessary,
it is straightforward to give back the dependence on $\mu$ and $\Mpl$.
In addition, 
as $\sigma_0$ corresponds to the freedom of the rescaling of the time coordinate,
without loss of generality, we may also set $\sigma_0=1$.
Therefore, 
only the remaining physical parameters are $f_0$, $\lambda_1$, and $\lambda_2$.

Proca star solutions 
have been constructed for the self-interacting potential with 
$\lambda_1\neq 0$ and $\lambda_2=0$
in Refs. \cite{Minamitsuji:2018kof,Herdeiro:2020jzx,Cardoso:2021ehg}.
Instead, 
in this work,
we will focus on the interaction $Z-Y^2$,
which vanishes in the case of the real Proca field,
and hence consider the case of 
\be
\label{choiceoflambdaa}
\lambda_1=0,\qquad \lambda_2\neq 0.
\ee
Clearly,
the most general case is that of $\left(\lambda_1\neq 0 ,\lambda_2\neq 0\right)$.
However, as far as we investigated several choices of $\left(\lambda_1\neq 0 ,\lambda_2\neq 0\right)$,
the physical properties of Proca stars such as the mass-radius relation
for the general case
can be simply understood by a combination of the two opposite cases,
i.e.,
$(\lambda_2\neq 0, \lambda_1=0)$
and 
 $(\lambda_1\neq 0, \lambda_2=0)$.
Thus, 
in this paper,
we focus on the case \eqref{choiceoflambdaa}
and its difference
from the opposite case  $(\lambda_1\neq 0, \lambda_2=0)$
discussed in Refs.~\cite{Minamitsuji:2018kof,Herdeiro:2020jzx,Cardoso:2021ehg}.

In Fig. \ref{fig_profile},
the profiles of 
the metric function $m$ (top),
the metric function $\sigma$ (middle),
and the Proca field (bottom)
for Proca star solutions in the the ground state
are shown as the functions of $\mu r$
for $f_0=0.05$.
The top and middle panels represent
the profiles of 
$m(r)$ and $\sigma(r)$,
while 
the solid and dashed curves in the lower panel correspond to the profiles of $a_0(r)$ and $a_1(r)$.
The top and middle panels show that 
both the metric functions $m(r)$ and $\sigma(r)$
approach larger/smaller constant positive values 
in the large distance regime $r\mu\gg 1$
for positive/negative values of $\lambda_2$,
than those for $\lambda_2=0$.
The bottom panel shows that 
the profiles of the Proca field
are more broadened 
in the vicinity of the center of the star for positive values of $\lambda_2$
and 
more compressed
for negative values of $\lambda_2$.
We note that the profiles of ${\tilde \alpha}_0(r)$ and ${\tilde \alpha}_1(r)$,
which will be defined in Eq. \eqref{alpha01},
are almost identical to $a_0(r)$ and $a_1(r)$ in Fig.~\ref{fig_profile}.
As for Proca star solutions in other theories
\cite{Brito:2015pxa,Brihaye:2017inn,Garcia:2016ldc,Minamitsuji:2018kof,Herdeiro:2020jzx,Cardoso:2021ehg,Zhang:2021xxa},
the temporal component of the complex Proca field $a_0(r)$ 
in the ground state
crosses zero once
and hence has a single node,
which is in contrast with the case of scalar boson stars
in the ground state
where the radial profile of the complex scalar field has zero nodes.

\subsection{Effective metric for the Proca field perturbations}
\label{sec3b}

On top of  Proca star solutions,
we consider small perturbations of the complex Proca field.
As is well known, the propagations of the longitudinal modes of $A_{\mu}$ are modified by the self-interactions, 
and a ghost or gradient instability may appear around a nontrivial background configuration of $A_{\mu}$. 
In this subsection, we study a high-frequency limit of the perturbations around a nontrivial background 
and find the effective metric on which the perturbations propagate.
A ghost (or gradient) instability appears when the signature 
of the temporal (or radial) component
of the effective metric changes.

We perturb these equations around a nontrivial background of $A_{\mu}$ and pick up the highest derivative terms
which characterize the propagation of the Proca field in the high-frequency limit.
The background is denoted by the same symbol $A_{\mu}$,
while the perturbations are
given by $\delta A_{\mu}+\partial_{\mu}\pi$,
where $\pi$ is the St\"{u}eckelberg field representing the longitudinal modes. 
One can easily find that Eq.~\eqref{eomProca} is reduced to the standard Maxwell equation in the high-frequency limit,
\begin{align}
\partial^2 \delta A_{\mu}-\partial^{\nu}\partial_{\mu}\delta A_{\nu} + \cdots =0,
\end{align}
where $\cdots$ are the terms that are at most linear in derivatives of the perturbations. On the other hand, the perturbed equation of \eqref{eomdiv} takes the form
\begin{align}
\bm{K}^{\mu\nu}\partial_{\mu}\partial_{\nu} 
\begin{pmatrix}
\pi^1 \\ \pi^2 
\end{pmatrix}
+\cdots =0
\end{align}
where $\pi^1$ and $\pi^2$ are the real and imaginary parts of $\pi$, and the components of the $2\times 2$ matrix $\bm{K}^{\mu\nu}$ are given by
\begin{align}
(\bm{K}^{\mu\nu})_{11}&=\left(\frac{\mu^2}{2}+\frac{\lambda_1}{2}Y -\lambda_2 A^{2\alpha} A^2_{\alpha} \right) g^{\mu\nu} 
\nn
&+ \lambda_1 A^{1\mu} A^{1\mu}+\lambda_2 A^{2\mu}A^{2\nu}
\,, \\
(\bm{K}^{\mu\nu})_{12}&=(\bm{K}^{\mu\nu})_{21}
\nn
&= \lambda_1-\lambda_2A^{1(\mu}A^{2\nu)} + \lambda_2 (A^{1\alpha} A^2_{\alpha}) g^{\mu\nu}
\,, \\
(\bm{K}^{\mu\nu})_{22}&=\left(\frac{\mu^2}{2}+\frac{\lambda_1}{2}Y -\lambda_2 A^{1\alpha} A^1_{\alpha} \right) g^{\mu\nu} 
\nn
&+ \lambda_1 A^{2\mu} A^{2\mu}+\lambda_2 A^{1\mu}A^{1\nu}
\end{align}
with $A_{\mu}=A^1_{\mu}+i A^2_{\mu}$. Here, the covariant derivatives are replaced with the partial ones because we are focusing on the short wavelength in comparison with the background spacetime curvature scale. The dispersion relation of the longitudinal modes is given by a root of
\begin{align}
{\rm det} (\bm{K}^{\mu\nu}k_{\mu}k_{\nu}) = 0
\label{dispersion_relation}
\end{align}
with $k_{\mu}$ being a four-wave-vector.

Either when $\lambda_1=0$ or $\lambda_2=0$, the dispersion relation \eqref{dispersion_relation} is factorized,
\begin{align}
{\rm det} (\bm{K}^{\mu\nu}k_{\mu}k_{\nu}) \propto k^2 \times g_{\rm eff}^{\mu\nu}k_{\mu}k_{\nu}=0
\quad ({\rm when}~\lambda_1 \lambda_2=0)
\,;
\end{align}
that is, one of the longitudinal modes propagates on the light cone of the spacetime metric,
while the other propagates on the effective metric $g^{\mu\nu}_{\rm eff}$.
In the following, let us concentrate on the case $\lambda_1=0$,
which is our main interest in the present paper. 
The effective metric with $\lambda_1=0$ is given by
\begin{align}
\label{eff_metric}
g^{\mu\nu}_{\rm eff}
&=g^{\mu\nu}\left[ (1-\lambda_2 Y/\mu^2)^2  - \lambda_2^2 Z/\mu^4 \right]
\nn
&+2A^{(\mu}\bar{A}^{\nu)}(\lambda_2/\mu^2 - \lambda_2^2 Y/\mu^4)
\nn
&+\frac{\lambda_2^2}{\mu^4}(A^{\alpha}A_{\alpha}\bar{A}^{\mu}\bar{A}^{\nu} + \bar{A}^{\alpha} \bar{A}_{\alpha} A^{\mu} A^{\nu})
\end{align}
of which the determinant is given by
\begin{align}
{\rm det}(g^{\mu\nu}_{\rm eff}) = {\rm det}(g^{\mu\nu}) \left[(1-\lambda_2 Y/\mu^2)^2  - \lambda_2^2 Z/\mu^4 \right]^3
\,.
\end{align}
In particular, the nonvanishing components of $\mathcal{H}^{\mu}{}_{\nu}:=g^{\mu\alpha}_{\rm eff}g_{\alpha\nu}$ under the spherically symmetric ansatz are
\begin{align}
\mathcal{H}^t{}_t &=  1 - 2\tilde{\lambda}_2 \tilde{\alpha}_1^2 
\label{htt}
\,, \\ 
\mathcal{H}^r{}_r &=  1 + 2\tilde{\lambda}_2 \tilde{\alpha}_0^2 
\label{hrr}
\,, \\
\mathcal{H}^{\theta}{}_{\theta}&=\mathcal{H}^{\varphi}{}_{\varphi}=(  1 - 2\tilde{\lambda}_2 \tilde{\alpha}_1^2  )(  1 + 2\tilde{\lambda}_2 \tilde{\alpha}_0^2  )
\,, 
\end{align}
where 
\begin{align}
\label{alpha01}
\tilde{\alpha}_0 := \frac{\tilde{a}_0}{\sigma(1-2m/r)^{1/2}}\,, \quad
\tilde{\alpha}_1 := \tilde{a}_1(1-2m/r)^{1/2}
\,.
\end{align}
A ghost or gradient instability
occurs at a point where ${\cal H}^t{}_t=0$ or ${\cal H}^r{}_r=0$,
respectively.
In the case of $\lambda_1=0$,
\be
{\cal H}(r)=r \mu^2 (r-2m)\sigma^2 {\cal H}^r{}_r,
\ee
where ${\cal H}(r)$ is defined in Eq. \eqref{calH}.
Thus, 
a gradient instability occurs
at the point ${\cal H}=0$,
where the background static solution to the equations \eqref{set_of_eqs}--\eqref{set_of_eqs4} ceases to exist.
Note that the condition to have a singular effective metric,
$(1-\lambda_2 Y/\mu^2)^2  - \lambda_2^2 Z/\mu^4 = \mathcal{H}^t{}_t \mathcal{H}^r{}_r=0$,
is expressed by coordinate-invariant scalar quantities,
meaning that the singularity is not a coordinate singularity.

In the case $\lambda_2<0$,
${\cal H}^t{}_t>0$ always,
but ${\cal H}^r{}_r$ may change the sign, and hence a gradient instability could occur.
As we have shown in Fig.~\ref{fig_profile}, the function $a_0$ takes the maximum value at $r=0$ and approaches zero as $r\to \infty$. The normalized function $\tilde{\alpha}_0$ also shows a qualitatively similar behavior.
Hence, 
${\cal H}^r{}_r$
 should take the minimum value at $r=0$ for $\lambda_2<0$ 
and whether 
${\cal H}^r{}_r$
crosses zero or not can be checked by the sign of the central value of 
${\cal H}^r{}_r$.
Using Eqs. \eqref{bc1}--\eqref{bc4},
in the vicinity of the center $r=0$,
$\mathcal{H}^r{}_r$
can be expanded as
\be
{\cal H}^r{}_r(r\to 0)
\to 1+\frac{2\lambda_2f_0^2}{\mu^2\sigma_0^2}.
\ee
Therefore, ${\cal H}^r{}_r$ does not cross zero
for
\be
f_0<\frac{\mu\sigma_0}{\sqrt{2|\lambda_2|}}\,,
\label{f0_up}
\ee
while ${\cal H}^r{}_r$ changes the sign if the central amplitude $f_0$ exceeds the critical value.
In other words, 
Proca star solutions can be constructed numerically
only for $f_0$ satisfying Eq. \eqref{f0_up}. 
If one wants to find solutions beyond the critical point,
one needs to include the effects of UV physics~\cite{Aoki:2022woy}.

In the case $\lambda_2>0$,
${\cal H}^r{}_r>0$ always, but ${\cal H}^t{}_t$ may change the sign,
and hence a ghost instability could occur.
As we will see later, 
for $\lambda_2>0$
a ghost instability appears for a sufficiently large central amplitude.
However, 
the point at which a ghost instability appears
is neither the vicinity of the center nor the large distance regime
because $\tilde{a}_1^2$ (or $\tilde{\alpha}_1^2$) takes the maximum value at a finite distance as shown in Fig.~\ref{fig_profile}.
The critical amplitude cannot be estimated analytically, and the critical amplitude will be obtained numerically in the next section.

\section{Properties of Proca stars}
\label{sec4}

In this section, we construct Proca star solutions numerically.
As mentioned in Sec. \ref{sec3},
for the numerical analysis we set $\Mpl=\mu=1$ and $\sigma_0=1$.
After constructing the solutions, 
we shall give back the dependence on $\Mpl$ and $\mu$ properly.

Since typically $\mu$ is assumed to be much smaller than
the Planck mass $\Mpl$,
the values of $|{\tilde \lambda}_i|$  ($i=1,2$)
given by Eq.~\eqref{dimless}
can be much larger than unity.
Hence, 
$|{\tilde \lambda}_i|={\cal O}(1)$ corresponds to the case with the Planck suppressed self-interactions,
and $|{\tilde \lambda}_i| \gg {\cal O}(1)$ may be interpreted as the case when the cutoff of the EFT is below $\Mpl$
(but still above $\mu$).
Note that in $\Mpl=1$ and $\mu=1$, ${\tilde \lambda}_i=\lambda_i$.

\subsection{$M-\omega$ and $Q-\omega$ relations}
\label{sec4a}

\subsubsection{$\lambda_2<0$ (${\tilde \lambda}_2<0$)}
\begin{figure}[h]
\unitlength=1.1mm
\begin{center}
  \includegraphics[height=4.5cm,angle=0]{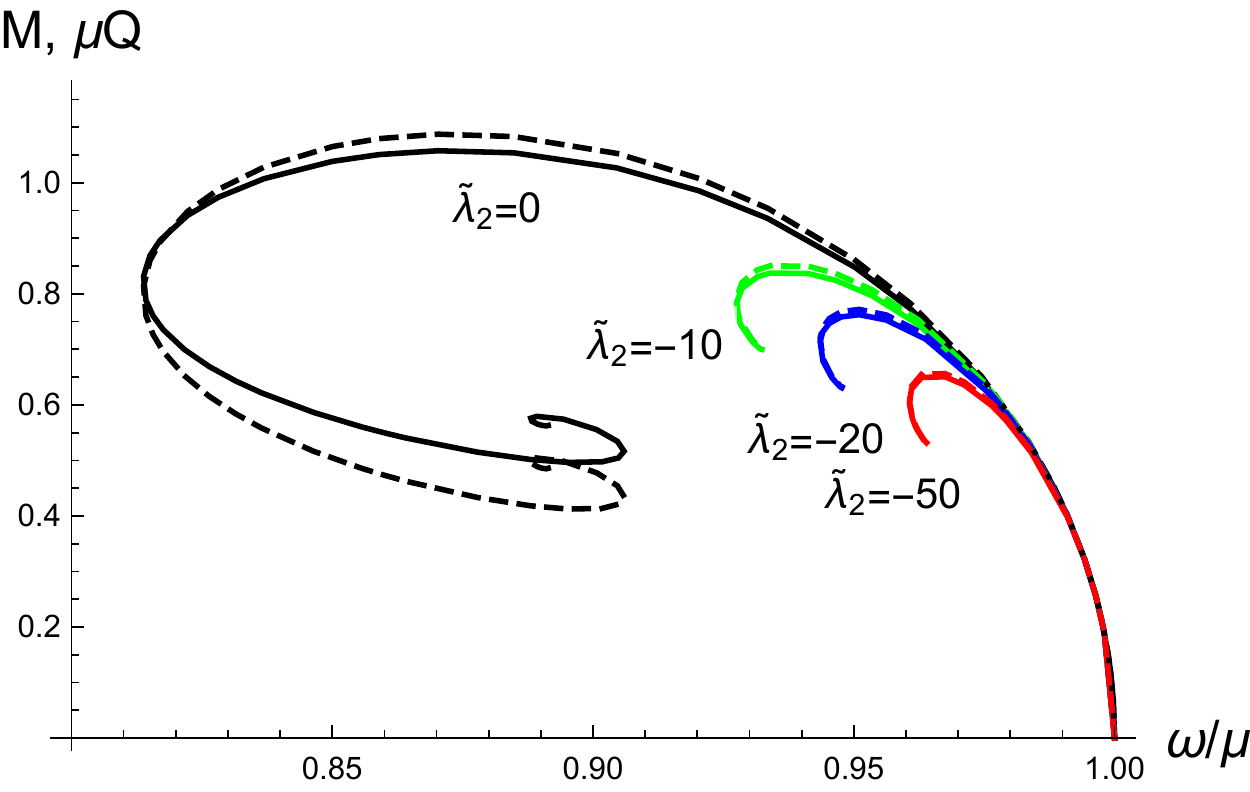}
\caption{
$M$ (solid curves) and $\mu Q$ (dashed curves) 
are shown as the functions of $\omega/\mu$ 
for ${\tilde \lambda}_2=0$ (black), $-10$ (green), $-20$ (blue), and $-50$ (red) from the top to the bottom.
$M$ and $\mu Q$ are shown in the units of $8\pi \Mpl^2/\mu$.
}
  \label{fig3}
\end{center}
\end{figure}

In Fig. \ref{fig3},
$M$ (solid curves) and $\mu Q$ (dashed curves) 
are shown as the functions of $\omega/\mu$ 
for ${\tilde \lambda}_2=0$ (black), $-10$ (green), $-20$ (blue), and $-50$ (red) from the top to the bottom.
$M$ and $\mu Q$ are shown in the units of $8\pi \Mpl^2/\mu$.
The curves for ${\tilde\lambda}_2<0$
are terminated at the amplitude saturating the bound \eqref{f0_up},
above which we cannot numerically construct Proca star solutions.
As argued in Ref.~\cite{Aoki:2022woy},
the onset of a gradient instability may not be an intrinsic pathology of the self-interacting Proca theories
but may be interpreted as the breakdown of the description of the Einstein-Proca theory \eqref{eft}
as a low-energy EFT.
For ${\tilde\lambda}_2<0$,
both $M$ and $\mu Q$ are suppressed,
and Proca star solutions exist only for the frequency $\omega$ relatively close to $\mu$.
Proca stars are gravitationally bound $\mu Q>M$.
Similarly to the conventional case ${\tilde\lambda}_2=0$,
the values of $M$ and $\mu Q$ do not differ so significantly.
The $M$-$\omega$ and $Q$-$\omega$ relations with ${\tilde \lambda}_2<0$ are qualitatively similar to the Proca stars with the self-interaction
${\tilde \lambda}_1<0$~\cite{Minamitsuji:2018kof}.

\subsubsection{$\lambda_2>0$ (${\tilde \lambda}_2>0$)}
\begin{figure}[h]
\unitlength=1.1mm
\begin{center}
  \includegraphics[height=4.5cm,angle=0]{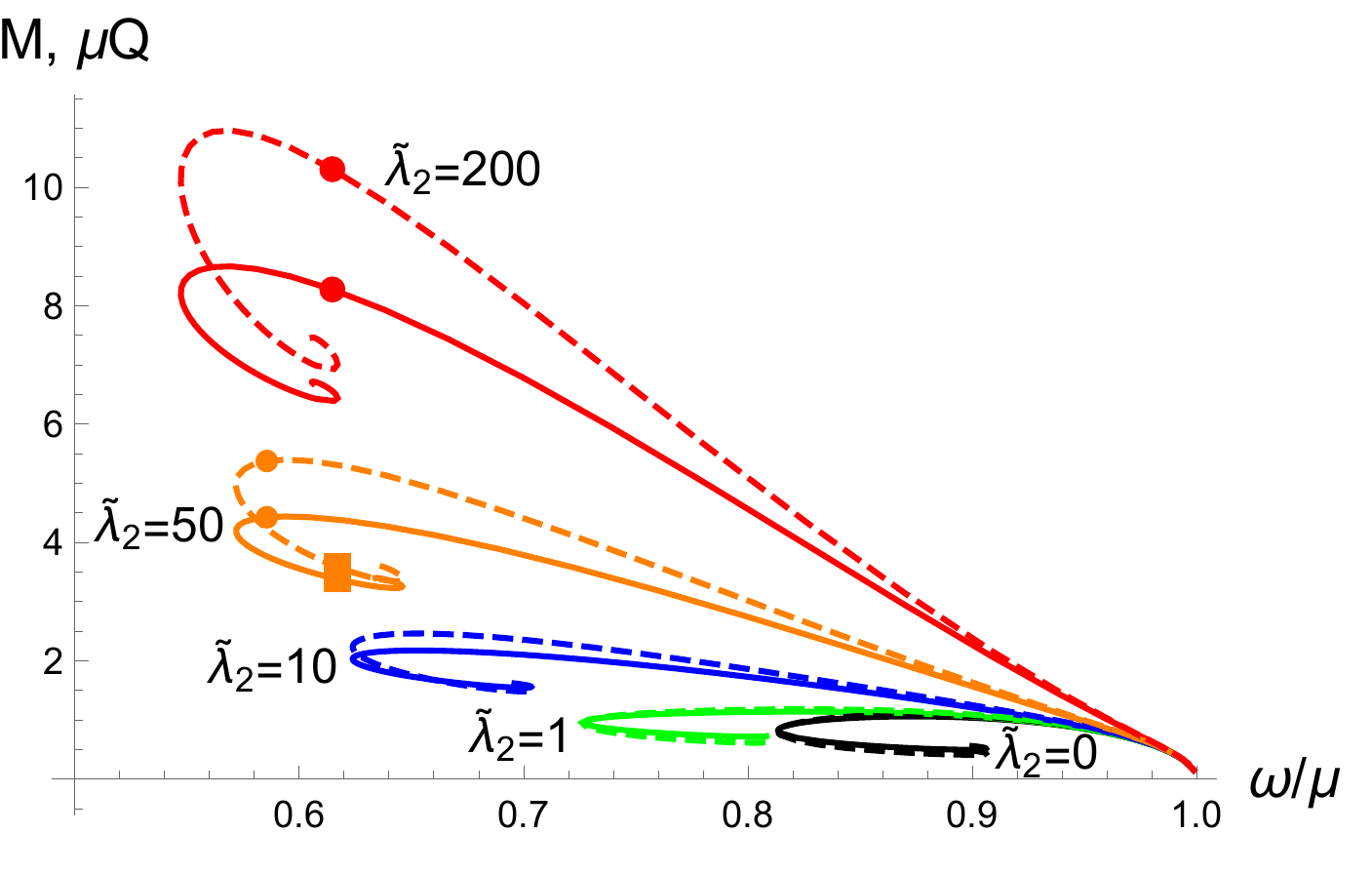}
\caption{
$M$ (solid curves) and $\mu Q$ (dashed curves) 
are shown as the functions of $\omega/\mu$
for 
${\tilde \lambda}_2=0$ (black), $1$ (green), $10$ (blue), $50$ (orange), and $200$ (red) from the bottom to the top.
$M$ and $\mu Q$ are shown in the units of $8\pi \Mpl^2/\mu$.
For $
{\tilde \lambda}_2=50$ and $200$, the points indicate the critical points beyond which a negative region of ${\cal H}^t{}_t$ appears (circles) and disappears (square).
}
  \label{fig2}
\end{center}
\end{figure}

Let us then consider the positive value of ${\tilde \lambda}_2$.
In Fig. \ref{fig2},
$M$ (solid curves) and $\mu Q$ (dashed curves) 
are shown as the functions of $\omega/\mu$
for ${\tilde \lambda}_2=0$ (black), $1$ (green), $10$ (blue), $50$ (orange), and $200$ (red)
from the bottom to the top.
$M$ and $\mu Q$ are shown in the units of $8\pi \Mpl^2/\mu$.
For ${\tilde \lambda}_2>0$,
we find that 
the self-interaction significantly enhances both $M$ and $\mu Q$.
However, unlike the conventional case, the value of $\mu Q$ is more enhanced than $M$, 
and then the $M$-$\omega$ relation is no longer close to the $Q$-$\omega$ relation for a large value of ${\tilde \lambda}_2$.
In particular, we find
that Eq.~\eqref{bound} is satisfied
even in the second branch of Proca stars, i.e.,~after reaching the minimum value of the frequency $\omega$, implying that Proca stars are gravitationally bound.

\begin{figure}[h]
\unitlength=1.1mm
\begin{center}
  \includegraphics[height=4.5cm,angle=0]{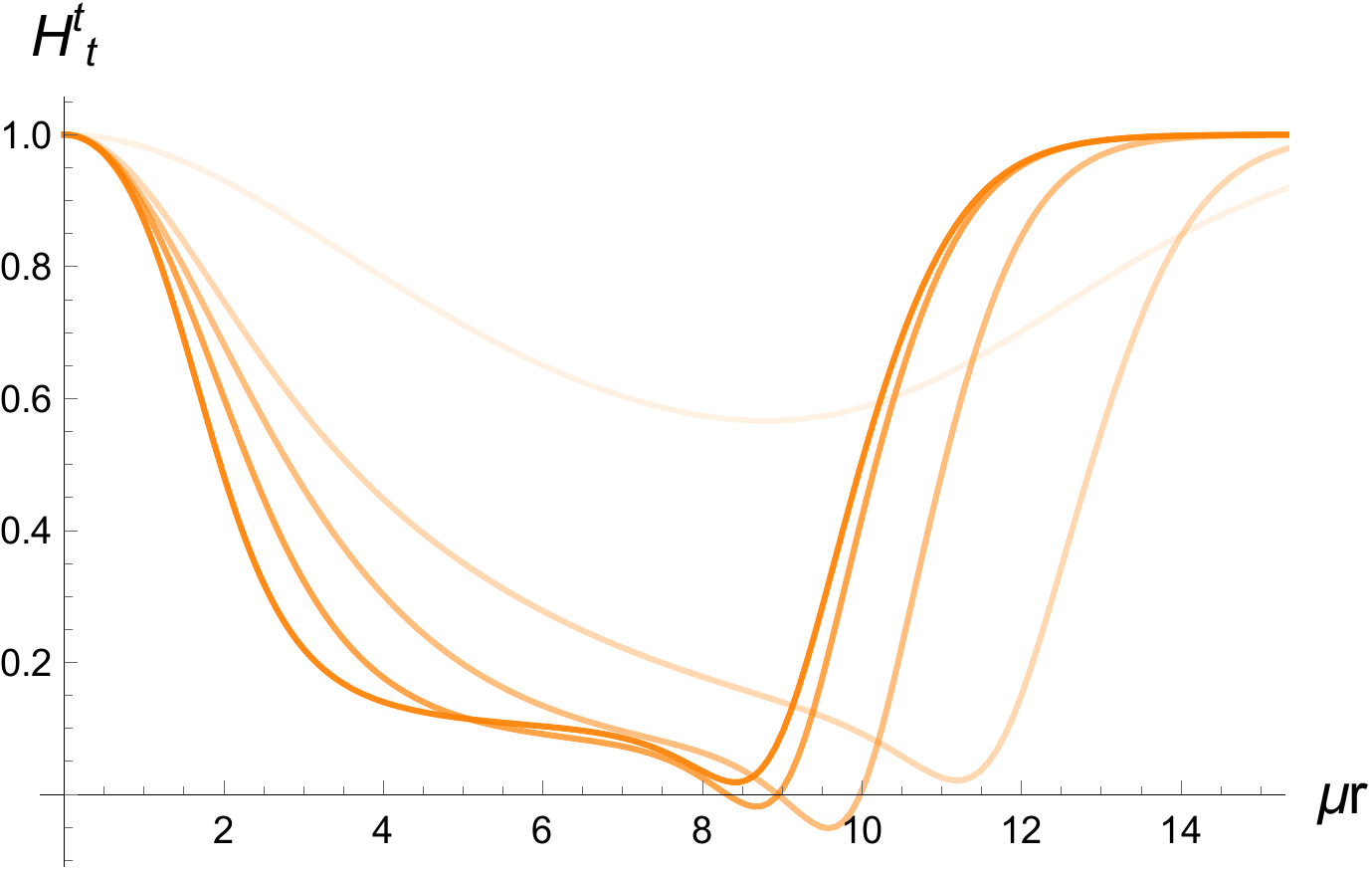}
\caption{
Profiles of the temporal component of the effective metric ${\cal H}^t{}_t$ for
${\tilde\lambda}_2=50$. A deeper color corresponds to a larger central amplitude of the Proca field.
}
  \label{figHtt}
\end{center}
\end{figure}

For ${\tilde \lambda}_2>{\cal O}(20)$, we find solutions that the temporal component of the effective metric changes the sign at an intermediate radius. We show several profiles of ${\cal H}^t{}_t$ with different central amplitudes of the vector field for $
{\tilde\lambda}_2=50$ in Fig.~\ref{figHtt}. As the central amplitude increases, the minimum value of ${\cal H}^t{}_t$ decreases, and nodes of ${\cal H}^t{}_t=0$ appear beyond a critical point. Proca star solutions then suffer from a ghost instability, which should invalidate the EFT description. On the other hand, the minimum value of ${\cal H}^t{}_t$ starts to increase if the central amplitude further increases, and then the nodes disappear. These critical points are shown in the circles and the squares in Fig.~\ref{fig2}, respectively. A negative region of ${\cal H}^t{}_t$ exists in solutions between the circle and the square. For ${\tilde\lambda}_2=200$, the profiles of ${\cal H}^t{}_t$ are qualitatively similar, and a negative region of ${\cal H}^t{}_t$ appears beyond a critical point as indicated by the circles in Fig.~\ref{fig2}. However, we numerically find that ${\cal H}^t{}_t$ still has a negative region,
even if the central amplitude further increases, differently from the case of ${\tilde \lambda}_2=50$.


\subsection{Mass-radius relations}
\label{sec4b}

\subsubsection{
$\lambda_2<0$
(${\tilde \lambda}_2<0$)
}

\begin{figure}[h]
\unitlength=1.1mm
\begin{center}
  \includegraphics[height=5.0cm,angle=0]{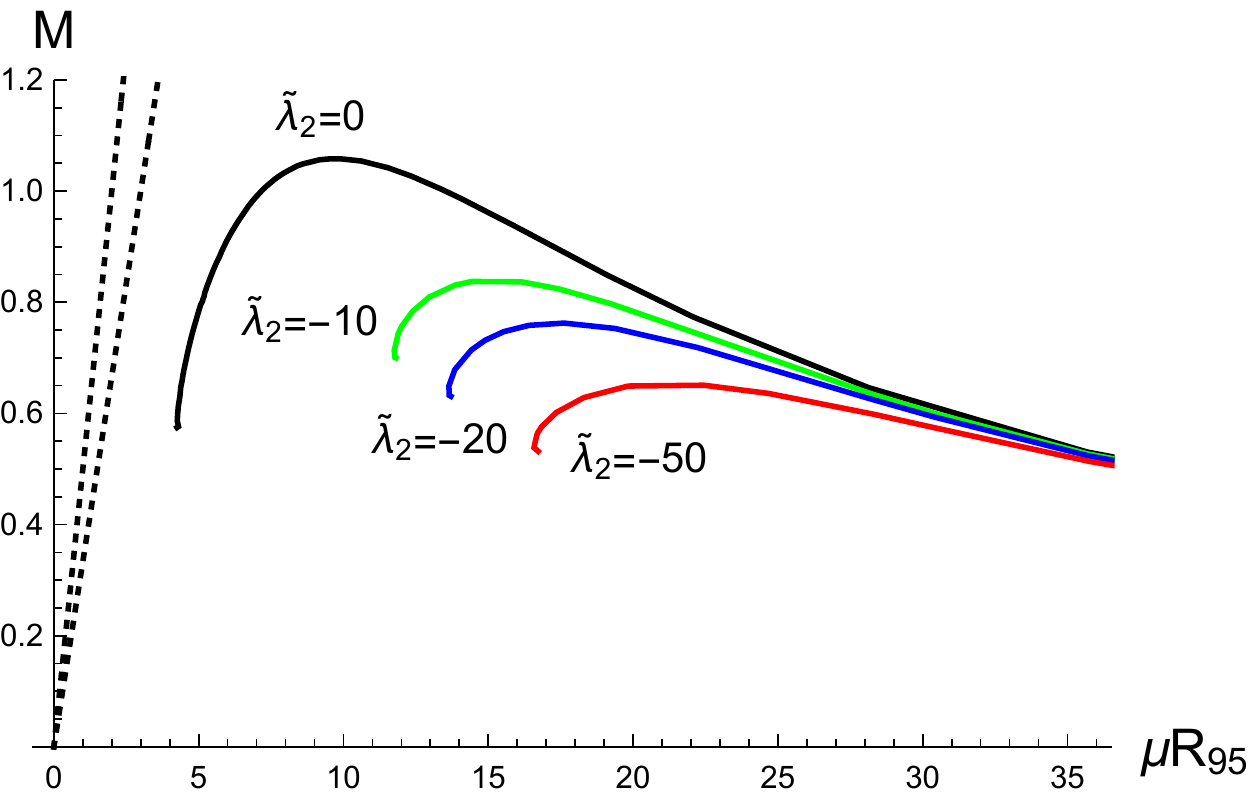}
  \includegraphics[height=5.0cm,angle=0]{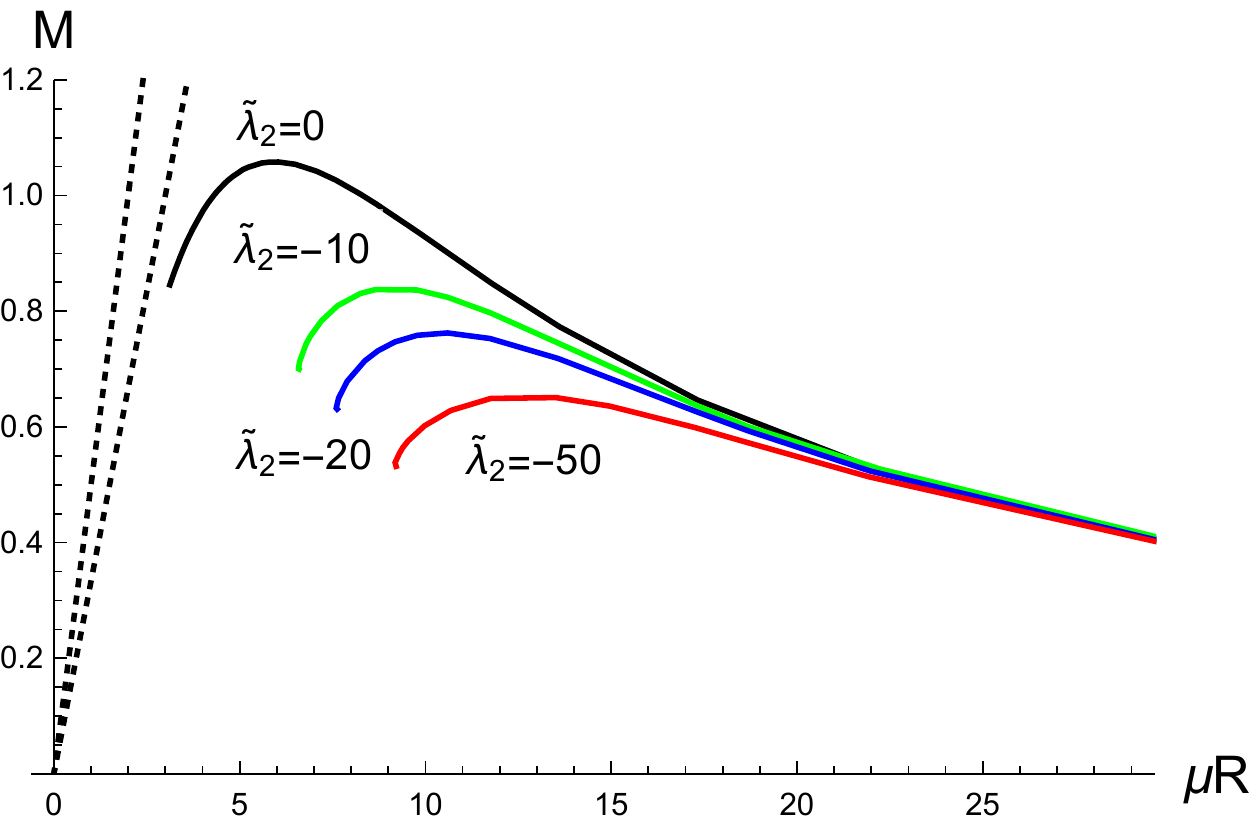}
\caption{
$M$ is shown as the function of  $\mu {\cal R}_{95}$ (top) and $\mu{\cal R}$ (bottom)
for ${\tilde \lambda}_2=0$
(black), $-10$ (green), $-20$ (blue), and $-50$ (red),
respectively.
The dotted lines from the top 
have the tilts $1/2$ and $1/3$, respectively.
$M$ is shown in the units of $8\pi \Mpl^2/\mu$.
}
  \label{fig6}
\end{center}
\end{figure}

In the top and bottom panels of Fig.~\ref{fig6},
$M$ is shown as the functions of  $\mu {\cal R}_{95}$ (top) and $\mu{\cal R}$ (bottom)
for ${\tilde \lambda}_2=0$ (black), $-10$ (green), $-20$ (blue), and $-50$ (red) [see Eqs.~\eqref{r95} and \eqref{reff}].
The dotted lines from the top 
have the tilts $1/2$ and $1/3$.
Crossing these lines
may indicate the formation of an event horizon and a photon sphere,
respectively.
As mentioned in Sec. \ref{sec3a},
although there is no unique definition of the surface of a Proca star,
${\cal R}_{95}$ should be more suitable to characterize it,
because ${\cal R}_{95}$ encloses the most of the volume where
the energy of the Proca field is contained,
especially for a Proca star with a more localized profile of the Proca field.
Thus, 
crossing the dashed line $M={\cal R}_{95}/3$ strongly indicates 
the formation of a photon sphere.
$M$ is shown in the units of $8\pi \Mpl^2/\mu$.
We find $\mR_{95}>\mR$ as mentioned previously.
The relative difference between $\mR_{95}$ and $\mR$
is about $60\%$ for the least compact stars
and 
exceeds $80\%$ for the most compact stars.
In comparison with the case of ${\tilde \lambda}_2=0$,
$M$ is always suppressed, and hence Proca stars become less compact
than the conventional case of ${\tilde \lambda}_2=0$.
Note that the other self-interaction $(\bar{A}^{\mu}A_{\mu})^2$ with ${\tilde \lambda}_1<0$ also leads to a similar effect, making Proca stars less compact.
In other words,
observationally,
it would be difficult to distinguish Proca star solutions for ${\tilde \lambda}_2<0$
from those in other theories.

\subsubsection{
$\lambda_2>0$ (${\tilde \lambda}_2>0$)
}
\begin{figure}[h]
\unitlength=1.1mm
\begin{center}
  \includegraphics[height=5.0cm,angle=0]{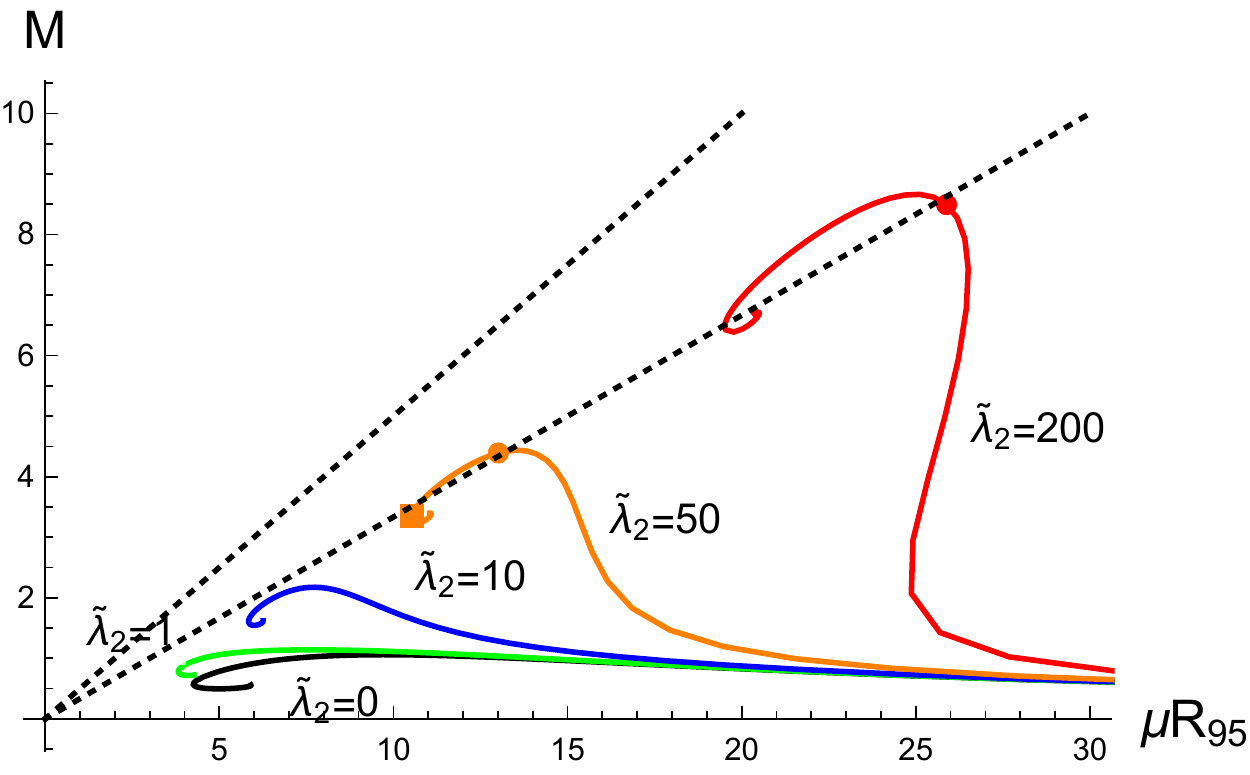}
  \includegraphics[height=5.0cm,angle=0]{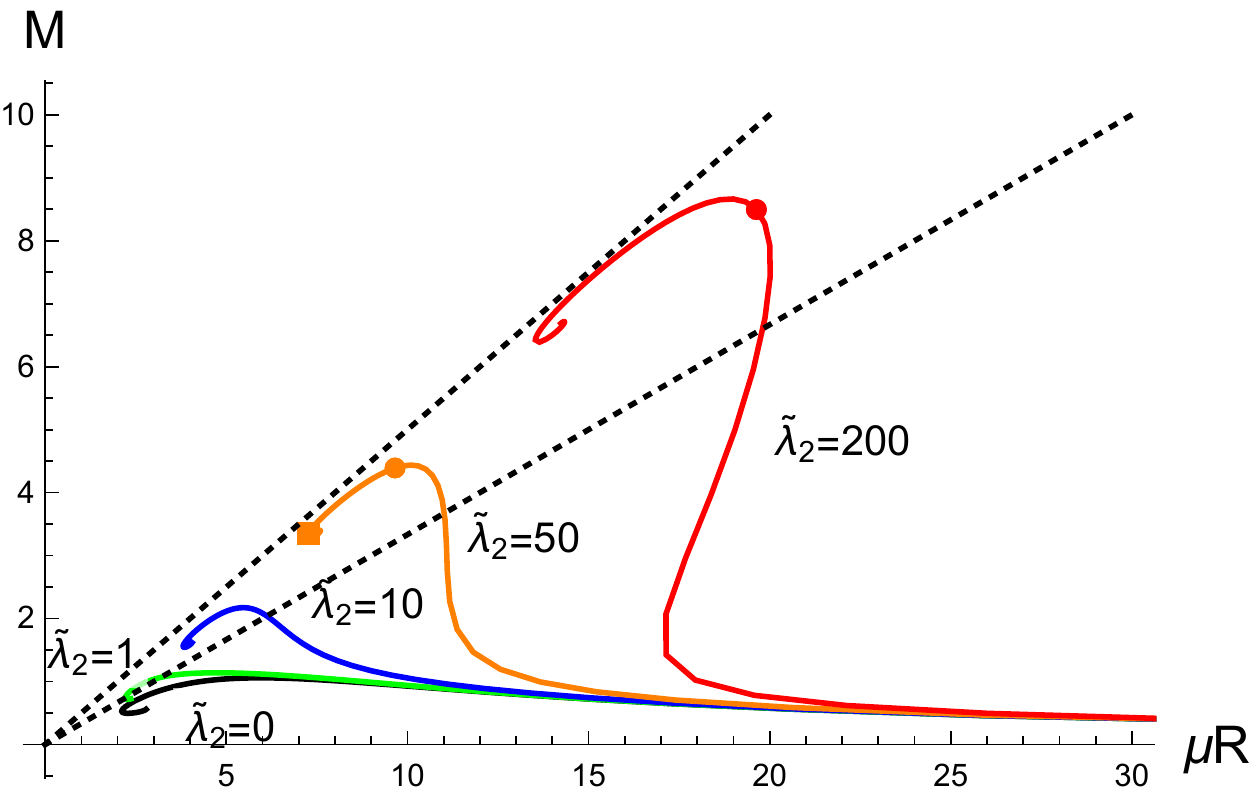}
\caption{
$M$ is shown as the function of $\mu {\cal R}_{95}$ (top) and $\mu {\cal R}$ (bottom)
for 
${\tilde \lambda}_2=0$ (black), $1$~(green), $10$~(blue), $50$~(orange), and $200$~(red).
The dotted lines from the top 
represent those with the tilts $1/2$ and $1/3$, respectively.
$M$ is shown in the units of $8\pi \Mpl^2/\mu$.
}
  \label{fig4}
\end{center}
\end{figure}

Opposite to the case with
${\tilde \lambda}_2<0$
the self-interaction with ${\tilde \lambda}_2>0$
gives compact Proca stars.
In Fig.~\ref{fig4},
$M$ is shown as the function of ${\cal R}_{95}$ (top) and ${\cal R}$ (bottom)
for ${\tilde \lambda}_2=0$ (black), $1$ (green), $10$ (blue), $50$ (orange), and $200$ (red) 
in the units of $8\pi \Mpl^2/\mu$.
The dotted lines from the top represent those with the tilts $1/2$ and $1/3$.
For a sufficiently large ${\tilde \lambda}_2$,
the mass-radius relation can exceed the line $M={\cal R}_{95}/3$, suggesting a formation of the photon sphere.
However, for ${\tilde \lambda}_2=50$ and $200$, the solutions with large amplitudes of the Proca field are suffered from a ghost instability, and the critical points appear in the vicinity of the line $M=\mR_{95}/3$. In particular, in the case of 
${\tilde \lambda}_2=50$,
both critical points (the circle and the square) are situated in the vicinity of the line $M=\mR_{95}/3$, and the solutions have a negative region of ${\cal H}^t{}_t$ only in $M\gtrsim {\cal R}_{95}/3$.
On the other hand, the curves for ${\tilde \lambda}_2=0,1,10$
 are free from a ghost instability but are always below the line $M=\mR_{95}/3$. Therefore, a ghost instability somehow forbids the formation of a photon sphere at least within the regime of validity of the EFT. We need a UV completion to properly discuss the properties of the Proca stars with $M\gtrsim {\cal R}_{95}/3$.

The bottom one of Fig.~\ref{fig4} shows the mass-radius relation by using the effective radius defined by \eqref{reff}
or the reference.
We always find that $\mR<\mR_{95}$,
and the curves
${\tilde \lambda}_2>0$ can
cross the line of $1/3$ even below the critical points.
The relative difference between $\mR_{95}$ and $\mR$
exceeds $60\%$ for the least compact stars
and 
is about $40\%$ for the most compact stars.
However, 
as mentioned previously,
the fact of $M>\mR/3$ may not indicate the formation
of the photon sphere,
because $\mR$ covers only a part of the region where the Proca field is localized, 
and should not be interpreted as an astrophysical surface of the star.

The difficulty for the formation of a photon sphere
has also been observed in the case of scalar boson stars \cite{Amaro-Seoane:2010pks,Giudice:2016zpa,Cardoso:2016oxy}.
For instance,
in the presence of the quartic-order self-interaction, the complex scalar field
$V_{\Phi}=\frac{\mu^2_{\Phi} }{2} \bar{\Phi} \Phi + \frac{\lambda_{\Phi} }{4} (\bar{\Phi} \Phi)^2$
[see Eq. \eqref{complex_scalar}],
and 
the maximal value of $M/{\cal R}_{\rm eff}$ for scalar boson stars with $\lambda_{\Phi}>0$
is about $0.16$,
while $M/{\cal R}_{\rm eff}$ for stars without the self-interaction $\lambda_\Phi=0$
is about $0.08$,
both of which are much below $1/3$ \cite{Amaro-Seoane:2010pks,Giudice:2016zpa}.
In Refs.~\cite{Amaro-Seoane:2010pks,Giudice:2016zpa},
 ${\cal R}_{99}$ analogous to ${\cal R}_{95}$ [see Eq.~\eqref{r95}]
was employed as an effective radius ${\cal R}_{\rm eff}$.
Note that in our case ${\cal R}_{99}/{\cal R}_{95}$
for the Proca star solutions on the verge of the onset of a ghost instability
is at most $1.10$ for ${\tilde \lambda}_2\approx 20$
and decreases as $\lambda_2$ increases,
so ${\cal R}_{95}$ differs only a little from ${\cal R}_{99}$ for highly compact stars of interest.
For the scalar potential $V_\Phi=\mu^2 |\Phi|^2(1-2|\Phi|^2/\sigma_0^2)^2$ 
with the sextic-order power of $|\Phi|$,
where $\sigma_0$ is a constant,
the scalar field has a very sharp profile,
and 
this self-interaction
allows us to realize the maximal value of $M/{\cal R}_{\rm eff}$ close to $1/3$ \cite{Cardoso:2016oxy}.
Thus, 
in the case of scalar boson stars,
although there is no ghost instability,
the quartic-order self-interaction 
is not strong enough to support highly compact boson stars.

On the other hand, 
in the case of Proca stars,
the quartic-order self-interaction~\eqref{quarticpotential} with Eq.~\eqref{choiceoflambdaa}
is strong enough to compress a Proca star.
However, 
the onset of a ghost instability 
indicates that
when one performs a numerical simulation 
of a collapse of a complex Proca field or a collision of two less compact Proca stars,
numerical evolution would break down 
before the formation of a highly compact single Proca star in the context of EFT,
as in other problems with a self-interacting Proca field~\cite{Clough:2022ygm,Coates:2022qia,Mou:2022hqb}.
Thus, 
unlike the case of scalar boson stars,
the formation of a photon sphere is prevented 
because of the breakdown of the time evolution of the Proca field.
It would be interesting to investigate whether 
the problem of a ghost instability could be cured by an inclusion of appropriate UV physics 
as discussed \cite{Aoki:2022woy}
(see also Ref. \cite{Barausse:2022rvg}).

Beyond the staticity and spherical symmetry,
it is known that
stationary and axisymmetric Proca stars without self-interactions can be compact enough
to have a light ring in the stable branch
(see, e.g., Refs.~\cite{Cunha:2022gde,Sanchis-Gual:2022mkk,Sengo:2022jif}).
It would be also interesting to 
explore rotating Proca star solutions in our theory \eqref{eft} with Eq.~\eqref{quarticpotential}
and investigate whether the ghost instability would prevent the formation of the photon surface.
For now, we leave these subjects for future studies.

\section{Conclusions}
\label{sec5}

In this paper,
we have investigated the properties of Proca star solutions 
in the Einstein-complex Proca theory 
with the quartic order self-interaction
$\lambda_2\left[\bar{A}^{\mu}\bar{A}_{\mu}A^{\nu}A_{\nu}-(A^\mu {\bar A}_\mu)^2\right]$,
in addition to the mass term $\mu^2 A^\mu \bar{A}_{\mu}$,
where $\lambda_2$ represents the dimensionless coupling constant.
Section \ref{sec2} was devoted to introducing our theory
from the comparison with the scalar-tensor theories.
This type of the quartic-order self-interaction
is different from $\lambda_1(A^\mu {\bar A}_\mu)^2$ previously considered,
where $\lambda_1$ also represents the dimensionless coupling constant,
and 
absent in the case of the real Proca field ${\bar A}_\mu=A_\mu$.
The coupling constants $\lambda_1$ and $\lambda_2$ have to be negative if the complex Proca theory has a standard UV completion, e.g.,~the Higgs mechanism. However, a positive $\lambda_1$ or $\lambda_2$ cannot be excluded in the gravitational setup (or if getting rid of one of the fundamental assumptions), although its UV completion is not known. In the present paper, we take a bottom-up approach and discuss the phenomenological consequences of the new self-interaction $\lambda_2$ with both positive and negative values.

In Sec. \ref{sec3},
we derived a closed set of the equations to determine the profile of Proca star solutions numerically.
These equations are of the first-order differential equations with respect to the radial coordinate $r$.
By solving them
in the vicinity of the center $r=0$,
the boundary conditions were obtained.
For an appropriate choice of the frequency of the complex Proca field,
we constructed Proca star solutions numerically
where the components of the Proca field are exponentially suppressed
while the metric variables exponentially approach constant in the large distance regions.
We focused on the ground state solutions 
where the temporal and radial profiles of a complex Proca field
have one and zero nodes, respectively.
As in the case of Proca star solutions with other potentials,
there are two conserved quantities which characterize Proca star solutions.
One is the ADM mass $M$ which can be read from the asymptotic value of the mass function,
and 
the other is the Noether charge $Q$ associated with the global $U(1)$ symmetry.
We also derived the effective metric for the propagation of the perturbation of the self-interacting Proca field
on top of a nontrivial Proca star background.
If the temporal and radial components of the effective metric vanish,
a Proca star solution suffers from ghost and gradient instabilities, respectively.
The analysis of the effective metric indicated that 
in the case $\lambda_2<0$ 
a Proca star solution could suffer from a gradient instability,
while in the case $\lambda_2>0$,
it could suffer from a ghost instability.

Section \ref{sec4}
was devoted to discussing the properties of the Proca star solutions.
For simplicity, we set $\lambda_1=0$
and focused on the case $\lambda_2\neq 0$.
The second type of the interaction is absent in the limit of the real Proca field and has not been explored in the context of the Proca star.
See Refs.~\cite{Minamitsuji:2018kof, Herdeiro:2020jzx, Cardoso:2021ehg} for the studies
about the first type of the self-interactions.
In the case of $\lambda_2<0$, we found that the ADM mass and Noether charge
are always suppressed compared to those in the case of the massive Proca theory. The role of $\lambda_2<0$ is qualitatively similar to the coupling $\lambda_1<0$. 
In both cases, a gradient instability occurs for a sufficiently large amplitude of the Proca field.
On the other hand, 
the properties of Proca stars for $\lambda_2>0$
are quite different from those for $\lambda_2<0$ (and those for $\lambda_1>0, \lambda_2=0$).
For $\lambda_2>0$,
the ADM mass and Noether charge 
are much more enhanced than those in 
the case of the purely massive Proca theory,
and the relative difference between them is significantly larger.
However, Proca star solutions with large compactness $M/{\cal R}_{95}\gtrsim 1/3$
suffer from a ghost instability at a finite radius.
Although for $\lambda_2>0$ Proca stars could also be significantly more compact
than those in the conventional case of the massive Proca theory,
we have found that the onset of a ghost instability practically forbids the formation of the photon sphere 
or requires a knowledge of the UV completion.
Although we did not explicitly discuss the most general case of $(\lambda_1\neq 0,\lambda_2\neq 0)$
in the main text,
we numerically investigated several examples
and
found that 
turning on both couplings does not give novel features of the Proca stars.
However, as we have not extensively studied all the parameter space, 
there might be an exceptional case which we leave for a future study.

In summary, if the Proca field is UV completed in a standard way
$(\lambda_1<0, \lambda_2<0)$, both self-interactions make Proca stars
less compact than those without self-interactions. 
On the other hand, the positive coupling constant $\lambda_2>0$ 
leads to compact Proca stars, which would be phenomenologically more interesting.
The properties of the Proca stars with $\lambda_2>0$ are qualitatively different from those of the conventional boson stars. It would be also remarkable that a ghost appears when the compactness exceeds $M/{\cal R}_{95}\simeq 1/3$, 
preventing the formation of a photon sphere in the regime of EFT.

It would be intriguing to see whether the pathology result of the formation of the photon sphere is generic in Proca stars. One may investigate Proca star solutions
in Einstein-complex Proca theories 
including higher-order powers of $Z(=A^\mu A^\nu {\bar A}_\mu {\bar A}_\nu)$
as well as $Y(={\bar A}^\mu A_\mu)$ 
[see Eqs.~\eqref{defy} and \eqref{defz}].
At any order power of $A_\mu$ and ${\bar A}_\mu$,
the self-interacting potential can be constructed
by a linear combination of powers of $Y$ and $Z$.
For instance, 
the most general sextic-order self-interacting potential 
is a linear combination of 
$Y^3$ 
and 
$YZ$,
and 
the most general octic-order one
is that of 
$Y^4$,
$Y^2Z$,
and 
$Z^2$.
The higher-order interactions become important for a sufficiently large Proca amplitude and may influence the properties of highly compact Proca stars. It would be interesting to explore the possibility of having a highly compact star before the onset of pathological instability thanks to the higher-order interactions. It is also important to see whether the pathology can be cured by appropriate UV physics as the ghost or gradient instability is expected to appear generically in the self-interacting Proca field.

\section*{ACKNOWLEDGMENTS}
K.A. would like to thank Waseda University for their hospitality during his visit.
The work of K.A. was supported in part by Grants-in-Aid from the Scientific Research Fund of the Japan Society for the Promotion of Science, No.~20K14468.
M.M.~was supported by the Portuguese national fund through the Funda\c{c}\~{a}o para a Ci\^encia e a Tecnologia in the scope of the framework of the Decree-Law 57/2016 of August 29, changed by Law 57/2017 of July 19, and the Centro de Astrof\'{\i}sica e Gravita\c c\~ao through the Project~No.~UIDB/00099/2020.

\appendix
\section{Positivity bounds on spin-1 fields}
\label{sec:positivity}
The positivity bounds are S-matrix constraints on the EFT to admit unitary, Poincar\'{e} invariant, causal, and local UV completion~\cite{Adams:2006sv}. Let us consider an $s\leftrightarrow u$ symmetric scattering amplitude ${\cal M}(s,t)$ of a nongravitational process $XY \to XY$ with fixed $t$ where $\{ s,t,u \}$ are the Mandelstam variables. The $s\leftrightarrow u$ crossing symmetry is manifest for a scattering of spin-0 particles, while special care is needed for spin-1 particles (and generic spinning particles)~\cite{Bellazzini:2016xrt,deRham:2017zjm}. However, in the forward limit $t\to 0$, which we shall focus on in the following, one can derive the positivity bounds for spin-1 particles similarly to spin-0 particles by working with linear polarizations~\cite{Bellazzini:2016xrt}. The mentioned properties lead to the so-called twice-subtracted dispersion relation which relates the amplitude with the integral of the imaginary part of the amplitude (and polynomial terms). In particular, the $s^2$ coefficient of the amplitude at $s=\mu_X^2+\mu_Y^2$ satisfies
\begin{align}
\frac{1}{2}\left. \frac{\D^2 {\cal M}(s,0)}{\D s^2} \right|_{s=\mu_X^2+\mu_Y^2}
= \frac{1}{\pi} \int_{\rm cut} \D s' \frac{ {\rm Im} {\cal M}(s', 0)}{(s'-\mu_X^2-\mu_Y^2)^3}
\,.
\label{s2coefficient}
\end{align} 
in the forward limit $t \to 0$, where the integral runs along the branch cut (and poles on the real axis if any), and $\mu_X$ and $\mu_Y$ are the masses of $X$ and $Y$, respectively. The left-hand side of \eqref{s2coefficient} is evaluated at low energy and can be therefore computed by the EFT. The right-hand side involves the integral in the high-energy region and cannot be explicitly computed unless UV completion is given. However, unitarity ensures the imaginary part of the forward limit amplitude is positive, leading to the bound
\begin{align}
\frac{1}{2}\left. \frac{\D^2 {\cal M}(s,0)}{\D s^2} \right|_{s=\mu_X^2+\mu_Y^2} > 0
\,.
\label{positivity_bd}
\end{align}
Therefore, the $s^2$ coefficient of the EFT amplitude has to be positive, which is known as the positivity bound.

When we decompose the complex Proca field into the real and imaginary parts, $A_{\mu}=A^1_{\mu}+iA^2_{\mu}$, the Proca field Lagrangian \eqref{proca_lagrangian} is given by
\begin{align}
\mathcal{L}&=\sum_{a=1}^2\left[ -\frac{1}{4}F^{a \mu\nu}F^a_{\mu\nu} - \frac{\mu^2}{2}A^{a\mu}A^a_{\mu} \right]
\nn
&-\frac{\lambda_1}{4}(A^{1\mu}A^1_{\mu}+A^{2\mu}A^2_{\mu})^2
\nn
&-\lambda_2[ (A^{1\mu}A^2_{\mu})^2 - A^{1\mu}A^1_{\mu} A^{2\mu}A^2_{\mu}]
\,.
\end{align}
The linear polarization basis of the spin-1 particle for the four-momentum $p^{\mu}=(p^0, 0, 0, p^3)$ is given by
\begin{align}
\epsilon_{\mu}^{T_1}=(0,1,0,0),~ 
\epsilon_{\mu}^{T_2}=(0,0,1,0), ~
\epsilon_{\mu}^L=\frac{1}{\mu}(p^3,0,0, p^0),
\end{align}
where $\epsilon_{\mu}^{T_1},\epsilon_{\mu}^{T_2}$ are the basis for transverse modes,
while $\epsilon_{\mu}^L$ is the basis for longitudinal mode, respectively. Since the interactions that we are considering do not involve derivatives, the scatterings of the transverse modes do not lead to the $s^2=\mathcal{O}(p^4)$ term at the tree level. The $s^2$ coefficient arises from the interaction $(A^{a\mu} A^b_{\mu})^2$ in the scattering $ab \to ab$ of the longitudinal modes where $a,b=1,2$. The coupling constants $\lambda_1$ and $\lambda_2$ are only relevant to $a=b$ and $a\neq b$, respectively. We thus find the inequalities \eqref{positivity_lambda} by applying the bound \eqref{positivity_bd} to the processes $aa\to aa$ and $ab\to ab~(a\neq b)$ in the Proca field Lagrangian \eqref{proca_lagrangian}.

The extension of the positivity bounds to gravitational theories is not straightforward. The main difficulty arises from the presence of the $s^2/t$ term due to the graviton $t$-channel exchange which gives a singular contribution in the forward limit. The singular contribution is canceled under the assumption that the amplitudes exhibit an appropriate high-energy behavior as predicted by string theory, and then the gravitational positivity bounds can be derived~\cite{Tokuda:2020mlf}. However, the bounds contain an uncertainty of the order of $\Mpl^{-2}$ due to the lack of precise knowledge of quantum gravity. A small violation of the nongravitational positivity bound \eqref{positivity_bd} could be possible in the gravitational system (see also Refs.~\cite{Hamada:2018dde, Alberte:2020jsk, Herrero-Valea:2020wxz, Caron-Huot:2021rmr, Alberte:2021dnj, Herrero-Valea:2022lfd} for related discussions). Allowing the small violation, the gravitational positivity bounds read
\begin{align}
\lambda_1,\lambda_2 <  \mathcal{O}(1) \times \frac{\mu^4}{\Mpl^2 M^2}
\end{align}
in the Einstein-Proca theory~\eqref{eft}, where $M$ is a scale depending on the details of the high-energy behavior of the amplitude. If the scale $M$ is given by the IR physics $M=\mathcal{O}(\mu)$, the coupling constants $\lambda_1,\lambda_2$ can take sufficiently large positive values to affect properties of Proca stars. On the other hand, the positive values are practically forbidden if $M$ is related to a UV physics scale, e.g.,~the cutoff scale of the EFT or the quantum gravity scale.

\bibliography{ref}
\bibliographystyle{JHEP}

\end{document}